\documentclass{jfm}

\usepackage{natbib}

\usepackage{amsmath}
\usepackage{graphicx}
\usepackage{caption}
\usepackage{subcaption}
\usepackage{float}
\usepackage{multirow}
\usepackage{bm}

\begin{document}

\newtheorem{lemma}{Lemma}
\newtheorem{corollary}{Corollary}

\shorttitle{Modal decomposition of fluid-structure interaction} %for header on odd pages
\shortauthor{A. Goza, T. Colonius} %for header on even pages

\title{Modal decomposition of fluid-structure interaction with application to flag flapping}

\author
 {
Andres Goza, \aff{1}
 \corresp{\email{ajgoza@gmail.com}},
  Tim Colonius\aff{1}
    }

\affiliation
{
\aff{1}
Department of Engineering and Applied Science, California Institute of Technology, USA
}

\maketitle

\begin{abstract}
Modal decompositions such as proper orthogonal decomposition (POD), dynamic mode decomposition (DMD) and their variants are regularly used to educe physical mechanisms of nonlinear flow phenomena that cannot be easily understood through direct inspection. In fluid-structure interaction (FSI) systems, fluid motion is coupled to vibration and/or deformation of an immersed structure. Despite this coupling, data analysis is often performed using only fluid or structure variables, rather than incorporating both. This approach does not provide information about the manner in which fluid and structure modes are correlated. We present a framework for performing POD and DMD where the fluid and structure are treated together.  As part of this framework, we introduce a physically meaningful norm for FSI systems. We first use this combined fluid-structure formulation to identify correlated flow features and structural motions in limit-cycle flag flapping. We then investigate the transition from limit-cycle flapping to chaotic flapping, which can be initiated by increasing the flag mass. Our modal decomposition reveals that at the onset of chaos, the dominant flapping motion increases in amplitude and leads to a bluff-body wake instability. This new bluff-body mode interacts triadically with the dominant flapping motion to produce flapping at the non-integer harmonic frequencies previously reported by \citet{Connell2007}. While our formulation is presented for POD and DMD, there are natural extensions to other data-analysis techniques. 
\end{abstract}

\section{Introduction}

Modal decompositions such as proper orthogonal decomposition (POD) and dynamic mode decomposition (DMD) have been used to distill important physical mechanisms from data, and to develop reduced-order models for turbulent wall-bounded flows \citep{Berkooz1993}, flow past a cylinder \citep{Chen2012,Bagheri2013}, and a jet in cross-flow \citep{Rowley2009,Schmid2010}, to name a few examples.

These techniques were developed for flows involving (at most) stationary immersed surfaces, and have been applied less extensively to fluid-structure interaction (FSI) problems, where the fluid motion is coupled to deformation and/or vibration of an immersed structure. In this FSI setting, data analysis has, to our knowledge, only been applied to data of either the fluid or the structure independently of the other. The fluid-only approach has been used to study flow past a flexible membrane \citep{Schmid2010}, a cantilevered beam \citep{Cesur2014}, and an elastically-mounted cylinder undergoing vortex-induced vibration \citep{Blanchard2017}. The solid-only approach has been applied to fish swimming \citep{Bozkurttas2009,Tangorra2010} and flag flapping \citep{Michelin2008,Kim2013}. These approaches reveal significant flow or structure behavior, respectively, but do not yield driving mechanisms in the omitted quantity. This in turn leaves the correlation between fluid and structure behavior unknown.

We propose a framework for data analysis of FSI systems where the fluid and structure are treated together, which naturally allows correlation between the fluid and structure to inform the resulting modes of the fully-coupled system. As part of this formulation, we define a norm in terms of the total mechanical energy of the FSI system. This combined fluid-structure data-analysis procedure is then demonstrated on limit-cycle flapping and chaotic flapping of strictly two-dimensional flags. We show that the methodology is useful in extracting the mechanisms of FSI in these various regimes.

We focus here on proper orthogonal decomposition (POD) and dynamic mode decomposition (DMD) because of their widespread use and their expected suitability for the problems considered here. The limit-cycle case described in section \ref{sec:LC} is associated with one dominant frequency, and thus DMD is a natural candidate because of its localized harmonic nature \citep{Mezic2013}. POD is also expected to be suitable because of the near-harmonic decomposition it typically yields for limit-cycle flows (such as occurs in vortex shedding past a cylinder near the critical Reynolds number of approximately 47; see, \emph{e.g.}, \citet{Kutz2016}). For the chaotic flapping problem described in section \ref{sec:chaos}, the non-broadband (`peaky') nature of the dynamics again makes DMD a fitting technique. However, POD and DMD are not ideal for all contexts. For example, \citet{Towne2017} demonstrated that in statistically stationary flows with broadband frequency content -- as observed in the majority of turbulent flows -- spectral POD provides an optimal decomposition. The major goal of the current work is to demonstrate the utility of performing data analysis in a manner that accounts for both the fluid and the structure, rather than explore the advantages of any particular technique, a question which in any event depends on the specific FSI problem under consideration. Future work can readily incorporate the methodology presented here into the appropriate technique for the intended application.

\section{POD and DMD of fluid-structure interaction}

We consider snapshot-based methods applied to discrete data. The associated data matrices are assumed to be organized so that each column provides the state of the system at an instance in time and each row contains the time history of a specific state variable. For simplicity we present our formulation in a two-dimensional setting; the extension to three dimensions is straightforward.

We assume fluid data is given on a stationary Cartesian grid, $\Omega$, made up of $n_f$ points ($\Omega \subset \mathbb{R}^{n_f}$), and let the streamwise and transverse fluid velocities at the $i^{th}$ time instance, $t_i$, be $\textbf{u}_i,\textbf{v}_i \in \Omega$. Fluid data is often provided in this format by immersed boundary methods and experiments; some numerical methods use moving meshes at each time step that conform to the moving structure, and fluid data obtained from these methods would need to be interpolated onto a single stationary grid at each time instance to use the method we propose here. Note that for FSI problems with bodies of finite (non-negligible) thickness, there may be points on $\Omega$ that lie within the body. In this case, the corresponding velocities $\textbf{u}_i, \textbf{v}_i$ should be set to zero to avoid spurious contributions from these `fictitious-fluid' quantities.

We consider structural data provided in a Lagrangian setting, with the structural domain, $\Gamma$, comprised of $n_s$ points ($\Gamma$ depends on time). We let $\boldsymbol{\chi}_i, \boldsymbol{\eta}_i \in \Gamma$ denote the streamwise and transverse structural displacements from an undeformed reference configuration at the $i^{th}$ time instance, and $\boldsymbol{\xi}_i, \boldsymbol{\zeta}_i \in \Gamma$ be the corresponding structural velocities. We define the total state vector at $t_i$ as $\textbf{y}_i = [\textbf{u}_i, \textbf{v}_i, \boldsymbol{\chi}_i, \boldsymbol{\eta}_i, \boldsymbol{\xi}_i, \boldsymbol{\zeta}_i ] ^T\in\mathbb{R}^{2n_f + 4n_s}$, and define the data matrix, $\textbf{Y}\in\mathbb{R}^{n\times m}$ ($n = 2n_f + 4n_s$ is the size of the state and $m$ is the number of snapshots), as $\textbf{Y} = [ \textbf{y}_1, \dots, \textbf{y}_m]$.

POD modes are computed from the mean-subtracted data matrix,  $\tilde{\textbf{Y}}$, whose $i^{th}$ column is defined as $\tilde{\textbf{Y}}_i = \textbf{Y}_i - \boldsymbol{\mu}$, where $\boldsymbol{\mu} = 1/m \sum_{k = 1}^m \textbf{y}_k$ is the sample temporal mean of $\textbf{Y}$. For DMD, \cite{Chen2012} found that the use of $\tilde{\textbf{Y}}$ reduces DMD to a discrete Fourier transform in time, and that using \textbf{Y} allows for growth-rate information to be retained.  For this reason, DMD is performed on \textbf{Y} below.

\subsection{Proper orthogonal decomposition}

POD decomposes the data into orthogonal spatially uncorrelated modes that are ordered such that the leading $k$ modes $(k \le m)$ provide the most energetically dominant rank-$k$ representation of $\tilde{\textbf{Y}}$. This optimal representation is defined with respect to a norm, and we therefore select an inner product space whose induced norm yields the mechanical energy of the FSI system. Defining $\textbf{x}$ as an Eulerian spatial coordinate and $\textbf{s}$ as a Lagrangian variable that parameterizes the structure, and letting $\textbf{u}(\textbf{x},t) = [u(\textbf{x},t), v(\textbf{x},t)]^T$, $\boldsymbol\chi(\textbf{s},t) = [\chi(\textbf{s},t), \eta(\textbf{s},t)]^T$, and $\boldsymbol{\xi}(\textbf{s},t) = [\xi(\textbf{s},t), \zeta(\textbf{s},t)]^T$ be continuous analogues of the discrete variables defined earlier, the mechanical energy is
\begin{equation}
E(t) = \frac{\rho_f}{2}\int_\Omega |\textbf{u}(\textbf{x}, t)|^2 d\textbf{x} + \int_\Gamma \left[ \kappa(\boldsymbol{\chi}(\textbf{s},t)) + \frac{\rho_s}{2} \left|\boldsymbol{\xi}(\textbf{s},t) \right|^2 \right] d\textbf{s}
\label{eqn:TE_cont}
\end{equation}
where $\Omega$ and $\Gamma$ are continuous analogous of the discrete domains defined earlier. The terms corresponding to the fluid and structural velocities represent the kinetic energy in the system ($\rho_f$ and $\rho_s$ are the fluid and structure density, respectively) and $\kappa(\boldsymbol{\chi}(\textbf{s},t))$ is the potential energy within the structure (for deforming bodies this is the strain energy). The potential (strain) energy for flapping flags will be defined in the next section. Note that for bodies of finite thickness where there is fictitious fluid in $\Omega \cap \Gamma$, we again assume the fluid velocity is set to zero within $\Gamma$. This can equivalently be viewed as subtracting the fictitious fluid contribution, $\rho_f/2 \int_\Gamma  |\textbf{u}(\textbf{x}, t)|^2 \delta(\textbf{x} - \boldsymbol{\chi}(\textbf{s},t) )d\textbf{s}$, from the definition of energy above.

While there are a variety of definitions of energy one could use (so long as it is the induced norm of an inner-product space), the mechanical energy is a natural choice because it is nonincreasing in time and accounts for the transfer of energy between the fluid and structure apart from viscous dissipation in the fluid. That is, through a straightforward computation one can show that in the absence of body forces and under the assumption that the shear stress is negligible on the boundary of $\Omega$ (which occurs for sufficiently large $\Omega$), 
\begin{equation}
\frac{dE(t)}{dt} = -2 \mu \int_\Omega \left( \nabla \textbf{u} + (\nabla \textbf{u})^T \right) :  \left( \nabla \textbf{u} + (\nabla \textbf{u})^T \right) d\textbf{x} \le 0
\label{eqn:diss}
\end{equation}
where $\mu$ is the dynamic viscosity of the fluid. Note that we assumed there is no dissipation in the structure in arriving at (\ref{eqn:diss}). Including this term would modify (\ref{eqn:diss}) by a term that depends on the properties of the structure but in any case is nonpositive.

In the discrete setting of interest, the norm is defined as $||(\cdot)||_\textbf{W} \equiv ||\textbf{W} (\cdot) ||_2$, where $\textbf{W}$ is a weighting matrix defined as
\begin{equation}
\textbf{W} = \begin{bmatrix} \sqrt{\frac{\rho_f}{2}}\textbf{I}^{2n_f} & \textbf{0}  & \textbf{0} \\ \textbf{0} & \textbf{L} &\textbf{0}  \\ \textbf{0} &\textbf{0} &  \sqrt{\frac{\rho_s}{2}} \textbf{I}^{2n_s} \end{bmatrix}
\end{equation}
In this expression, $\textbf{I}^n$ is the $n\times n$ identity matrix and \textbf{L} is the operator that maps the structural displacements to the potential energy of the structure. We assume that \textbf{L} is formulated to be positive definite and symmetric so that \textbf{W} is positive definite and symmetric.

The inner product associated with this weighting matrix is defined as $\langle\textbf{q}, \textbf{p}\rangle_\textbf{W} \equiv \textbf{q}^T\textbf{W}^2\textbf{p} = (\textbf{W}\textbf{q})^T( \textbf{W}\textbf{p})$ $\forall \textbf{q},\textbf{p} \in \mathbb{R}^n$ and the induced norm is $||\textbf{q}||_\textbf{W} \equiv \sqrt{\langle \textbf{q}, \textbf{q} \rangle_\textbf{W}} = \sqrt{(\textbf{W}\textbf{q})^T( \textbf{W}\textbf{q})}$ $\forall \textbf{q} \in \mathbb{R}^n$, which is a discrete approximation of the square root of (\ref{eqn:TE_cont}) scaled by one on the length between data points, $\Delta x$. (This assumes that the distance between points of the fluid and structural domains is equal; unequal spacings can be incorporated into \textbf{W} in the standard ways).

The energetically ordered POD modes with respect to the \textbf{W}-weighted norm may be written in terms of the singular value decomposition (SVD) $\textbf{W}\tilde{\textbf{Y}} = \textbf{U}\boldsymbol{\Sigma}\textbf{V}^T$, where $\boldsymbol{\Sigma}$ is a diagonal matrix containing the singular values $\sigma_1, \dots, \sigma_m$ ordered by decreasing energy, and \textbf{U} (\textbf{V}) has columns $\textbf{u}_j$ ($\textbf{v}_j$) containing the left (right) singular vectors that correspond to $\sigma_j$. In this notation, the POD modes are $\hat{\textbf{U}} \equiv \textbf{W}^{-1} \textbf{U}$ (note that they are orthogonal with respect to the $\textbf{W}$-weighted inner product). These modes are written in terms of the SVD, but may be computed more efficiently using the method of snapshots \citep{Sirovich1987}. The energetically optimal rank-$k$ ($k\le m$) approximation of a snapshot $\textbf{y}_i$ may be expressed through an orthogonal projection onto the POD modes as
\begin{equation}
\textbf{y}_i \approx \sum_{j=1}^k \hat{\textbf{u}}_j^T (\textbf{W}\textbf{y}_i)\hat{\textbf{u}}_j
\label{eqn:POD_approx}
\end{equation}

\subsection{Dynamic mode decomposition}

Whereas POD modes define an energetically optimal representation of the data, DMD modes are obtained from a linear regression that best represents the dynamics of a (potentially nonlinear) data set. Though there are more general variants \citep{Tu2014}, we compute DMD modes from the matrix $\textbf{A}$ that best maps the progression of the state from one time instance to the next; \emph{i.e.}, the \textbf{A} that satisfies $\min \sum_{j=1}^{m-1} || \textbf{y}_{j+1} - \textbf{A}\textbf{y}_j ||_2$\footnote{The minimization can also be performed with respect to the \textbf{W}-weighted norm, but we retain the use of the standard 2-norm for consistency with most approaches in the literature.}. This relation can often be satisfied exactly under reasonable conditions on the data (such as linear independence of the columns of \textbf{Y}), and the best-fit matrix is $\textbf{A} = \textbf{Y}' (\textbf{Y}'')^{\#}$, where $\textbf{Y}' = [\textbf{y}_2, \dots, \textbf{y}_m]$, $\textbf{Y}'' = [\textbf{y}_1, \dots, \textbf{y}_{m-1}]$, and $(\textbf{Y}'')^\#$ is the pseudo-inverse of $\textbf{Y}''$. 

DMD modes are the eigenvectors of \textbf{A}, denoted as $\boldsymbol\Phi = [\boldsymbol\phi_1, \dots, \boldsymbol\phi_{m-1}]$. These modes may be computed efficiently without forming \textbf{A} explicitly \citep{Tu2014}.  The corresponding eigenvalues, $\hat{\gamma}_1, \dots, \hat{\gamma}_{m-1}$, are structured such that $\hat{\gamma}_j = e^{2\pi\gamma_j \Delta t}$, where $\Delta t$ is the time step between two snapshots and $\gamma_j$ is a complex number whose real and imaginary parts give the growth rate and frequency, respectively, of mode $j$. Note that $\gamma_j$ may be computed from $\hat{\gamma}_j$ via $\gamma_j =\log(\hat{\gamma}_j) / (2\pi\Delta t)$. A $k^{th}$ order ($k \le m-1$) representation of the system at the $i^{th}$ time instance $t_i$ may be written in terms of the DMD modes as
\begin{equation}
\textbf{y}_i \approx \sum_{j=1}^k c_j e^{2\pi\gamma_j t_i} \boldsymbol\phi_j
\label{eqn:DMD_approx}
\end{equation}
where $c_j = (\boldsymbol\Phi^\# \textbf{y}_1)_j$ represents the initial condition in terms of the $j^{th}$ DMD mode.

The above describes the DMD formulation derived for flows without bodies or flows involving stationary bodies, and may be used without modification for FSI problems to obtain the coupled flow-structure behavior that best represents the full system dynamics.

\section{Application to flag flapping}
\label{sec:flags}

The dynamics of flag flapping are governed by the Reynolds number ($Re$) and the dimensionless mass ($M_\rho$) and bending stiffness ($K_B$), defined as
\begin{equation}
Re = \frac{\rho_f U L}{\mu}, \; M_\rho = \frac{\rho_s h}{\rho_f L}, \; K_B = \frac{EI}{\rho_f U^2 L^3}
\end{equation}
where $\rho_f$ ($\rho_s$) is the fluid (structure) density, $U$ is the freestream velocity, $L$ is the flag length, $\mu$ is the dynamic viscosity of the fluid, $h$ is the flag thickness, and $EI$ is the bending stiffness. 

The potential (strain) energy in the flag is given by the flag displacement in the direction normal to the flag, $\chi_n(s,t)$, as $\kappa(\chi_n(s,t)) = K_B\partial^2 \chi_n / \partial s^2$ (note that for flags only one Lagrangian variable is required to parametrize the body, so the scalar $s$ is used in place of $\textbf{s}$). In the case of inextensible flags considered here, the strain energy may be written in terms of the streamwise and transverse displacements as $\kappa(\boldsymbol\chi(s,t)) = K_B(\partial^2 \chi/ \partial s^2 + \partial^2 \eta / \partial s^2)$. We therefore define the $\textbf{L}$-submatrix of \textbf{W} using the standard second-order central difference formula for the $\chi$ and $\eta$ sub-blocks, which results in a symmetric positive definite weighting matrix.

 The data for this analysis was obtained using the immersed-boundary method of \cite{Goza2017}. The method allows for arbitrarily large flag displacements and rotations, and is strongly-coupled to account for the nonlinear coupling between the flag and the fluid. The method was validated on several flapping flag problems. The physical parameters for each run are described in the subsequent subsections; see \cite{Goza2017} for details about the simulation parameters such as the grid spacing and time step that were used for the different simulations.

\subsection{Limit-cycle flapping}
\label{sec:LC}

\begin{figure}
\centering
	\begin{subfigure}[b]{0.3\textwidth}
		\hspace*{-15mm}
        		\includegraphics[scale=0.25,trim={0cm 0cm 0cm 0cm},clip]{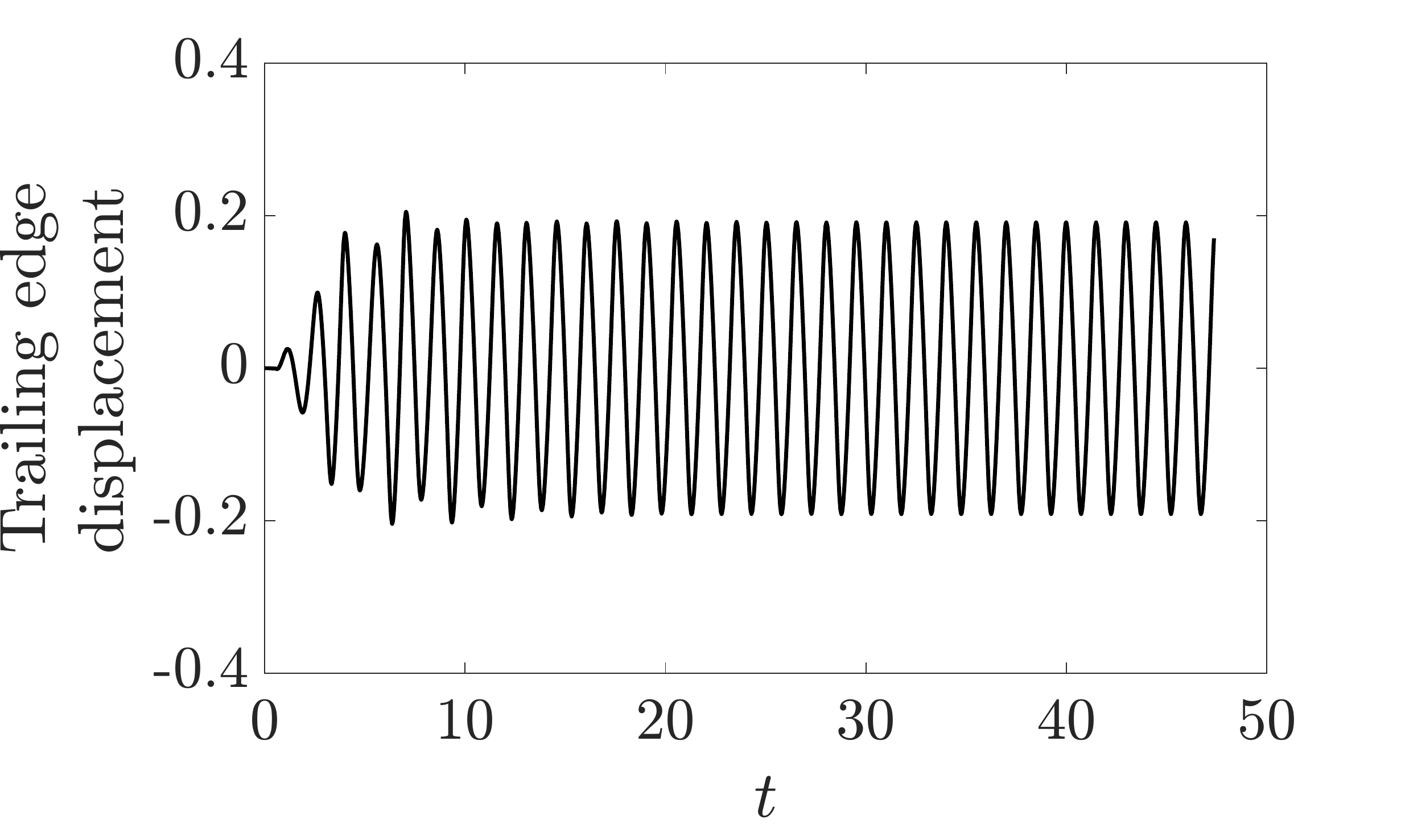}
	\end{subfigure}
	\begin{subfigure}[b]{0.3\textwidth}
		\hspace*{12mm}
        		\includegraphics[scale=0.25,trim={0cm 0cm 0cm 0cm},clip]{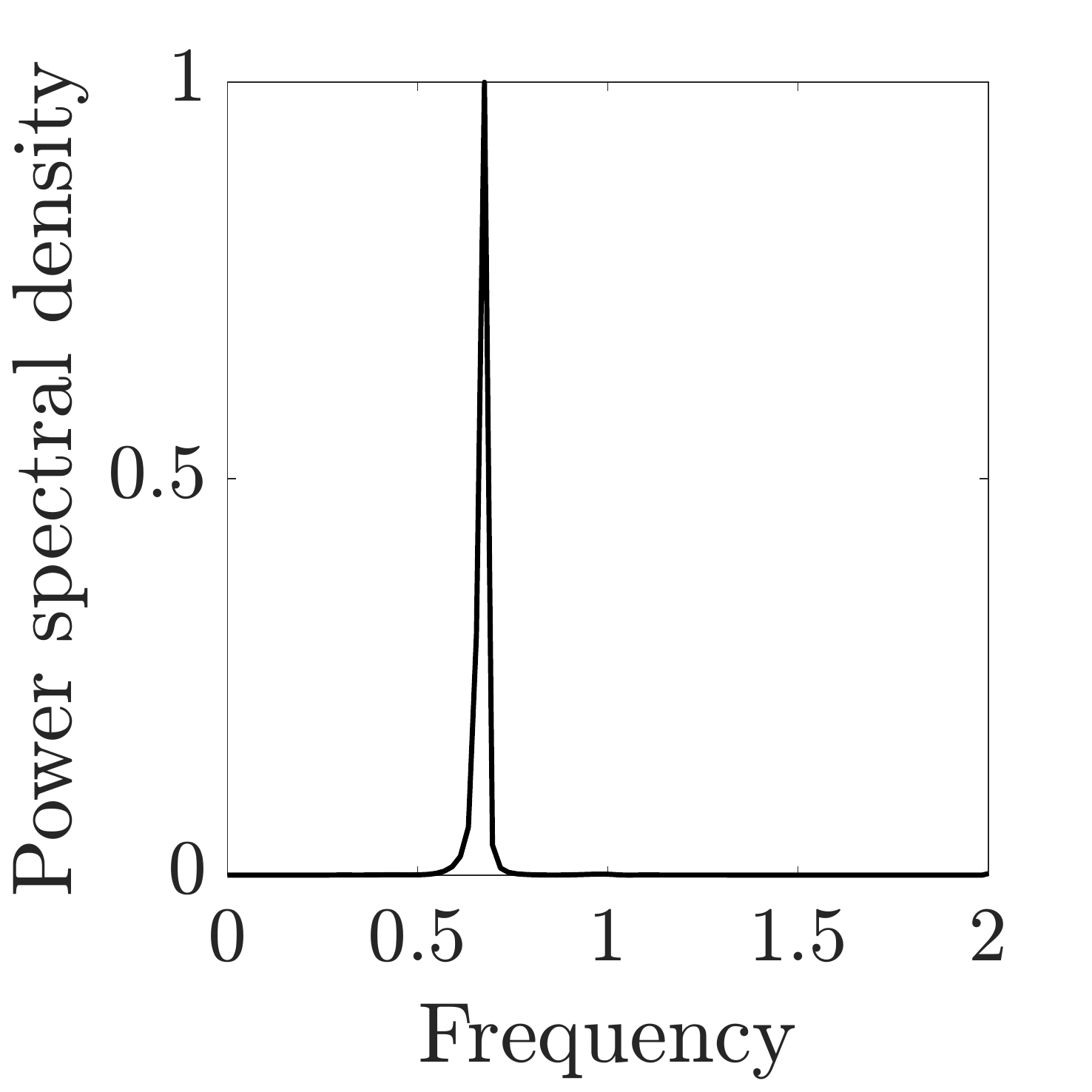}
	\end{subfigure}
	\caption{Transverse displacement (left) and spectral density (right) of the trailing edge of a flag in limit-cycle flapping with $Re =500$, $M_\rho = 0.18$, and $K_B = 0.0001$.}
	\label{fig:conv_LC_tip}
\end{figure}

\begin{figure}
	\begin{subfigure}[b]{0.245\textwidth}
        		\includegraphics[scale=0.35,trim={0cm 0cm 0cm 0cm},clip]{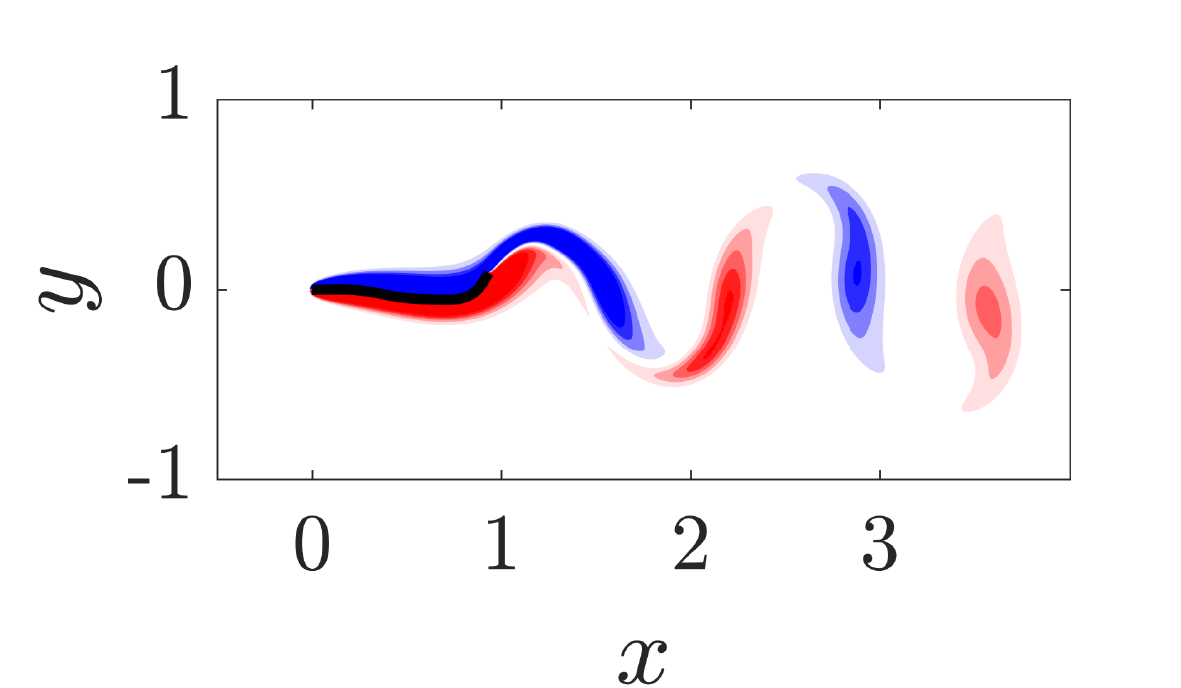}
	\end{subfigure}
	\begin{subfigure}[b]{0.245\textwidth}
		\hspace*{3.9mm}
        		\includegraphics[scale=0.35,trim={2cm 0cm 0cm 0cm},clip]{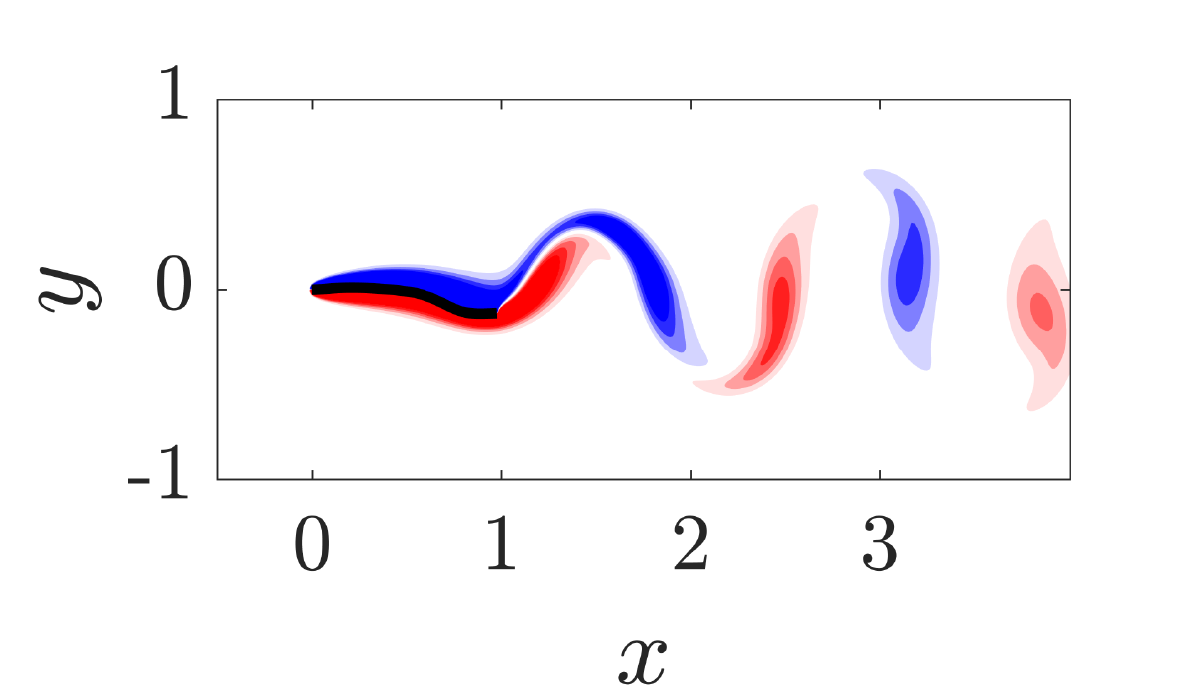}
	\end{subfigure}
    	\begin{subfigure}[b]{0.245\textwidth}
		\hspace*{1.5mm}
        		\includegraphics[scale=0.35,trim={2cm 0cm 0cm 0cm},clip]{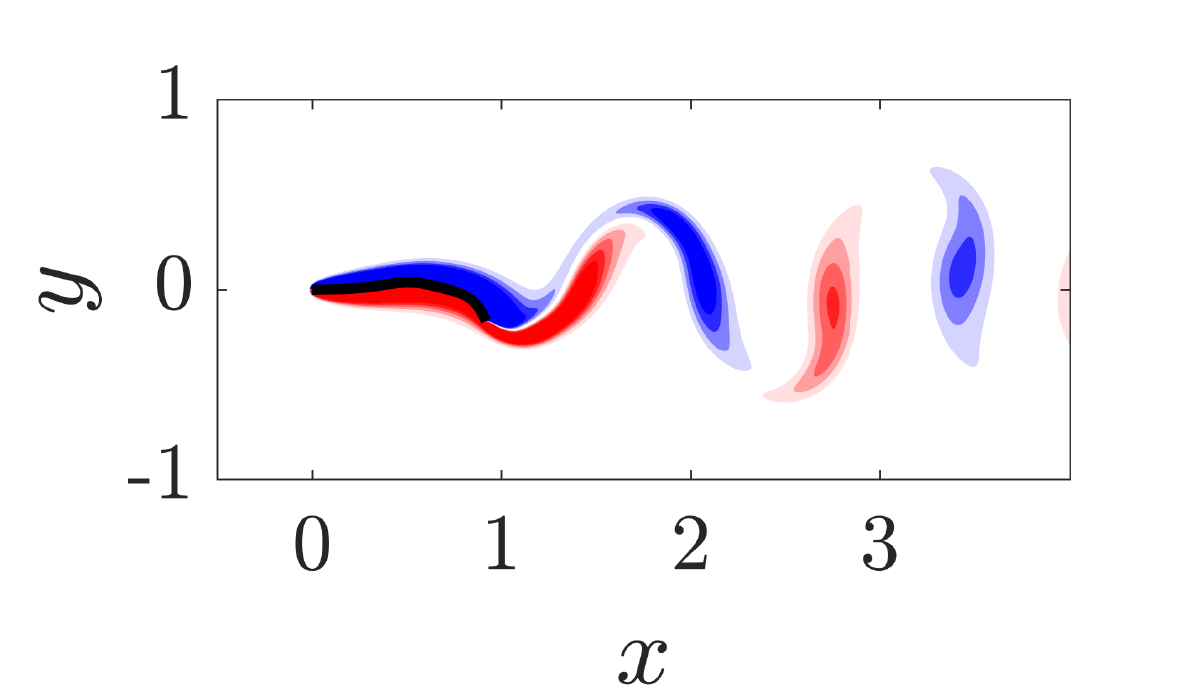}
	\end{subfigure}
	\begin{subfigure}[b]{0.245\textwidth}
		\hspace*{-0.8mm}
        		\includegraphics[scale=0.35,trim={2cm 0cm 0cm 0cm},clip]{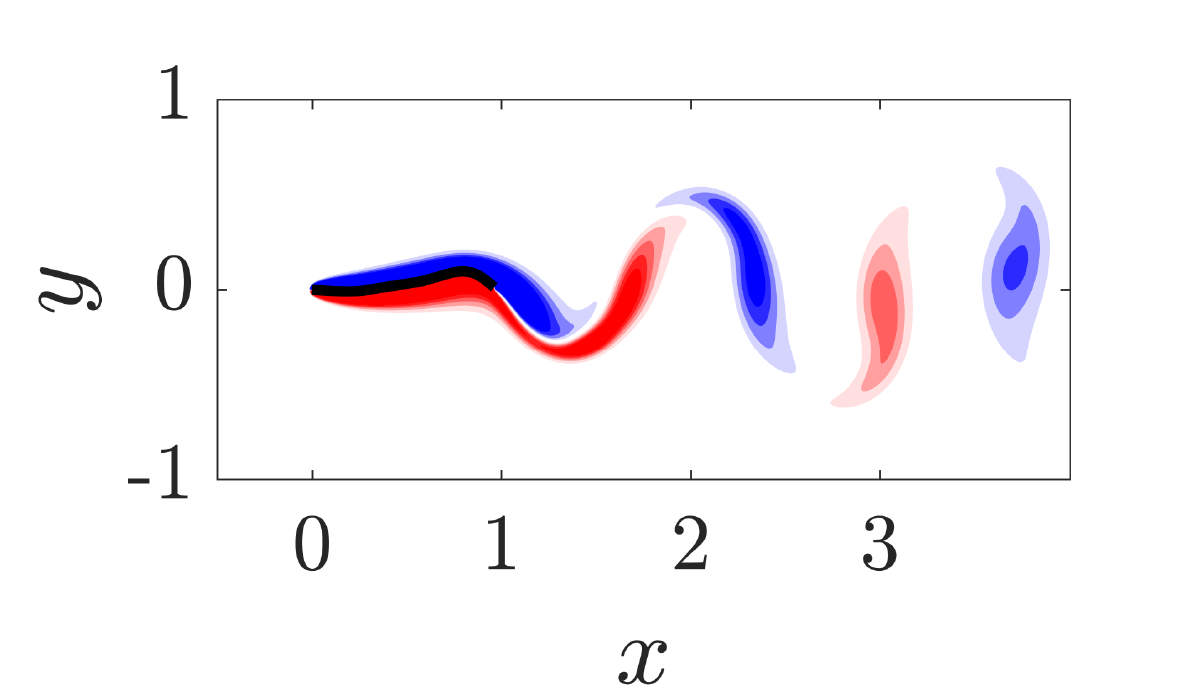}
	\end{subfigure}
    \caption{Snapshots of a flapping period for a flag in limit-cycle flapping with $Re =500, M_\rho = 0.18, K_B = 0.0001$. Contours are of vorticity, in 18 increments from -5 to 5.}
	\label{fig:LC_conventional}
\end{figure}

We consider a POD and DMD analysis of flapping with $Re =500, M_\rho = 0.1,$ and $K_B = 0.0001$, for which the system enters limit-cycle behavior \citep{Connell2007}. Figure \ref{fig:conv_LC_tip} shows the transverse displacement of the trailing edge of the flag as a function of time along with the corresponding power spectral density. Our analysis is performed after the transient region, once the system enters periodic behavior of fixed amplitude and frequency (beginning at $t \approx 20$ in figure \ref{fig:conv_LC_tip}). Figure \ref{fig:LC_conventional} shows contours of vorticity at four snapshots in time during a period of flapping in the limit cycle regime.  Snapshots were obtained over the range $t \in [20,40]$ in increments of $\Delta t =0.05$.

\begin{figure}
\centering
	\begin{subfigure}[b]{0.45\textwidth}
%	\hspace*{15mm}
%	\vspace*{5mm}
            	\includegraphics[scale=0.45,trim={0cm 0cm 0cm 0cm},clip]{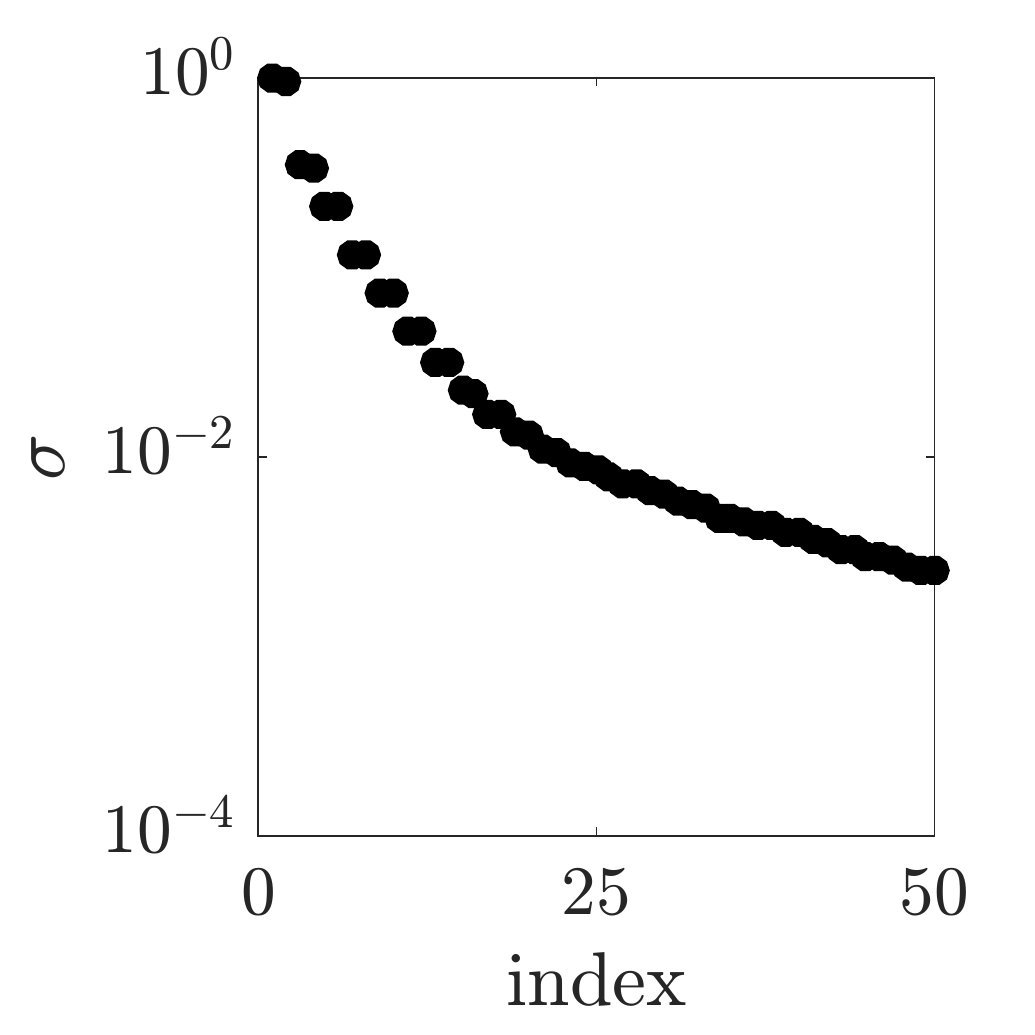}
	\end{subfigure}
	\begin{subfigure}[b]{0.45\textwidth}
	\hspace*{0mm}
            	\includegraphics[scale=0.45,trim={0cm 0cm 0cm 0cm},clip]{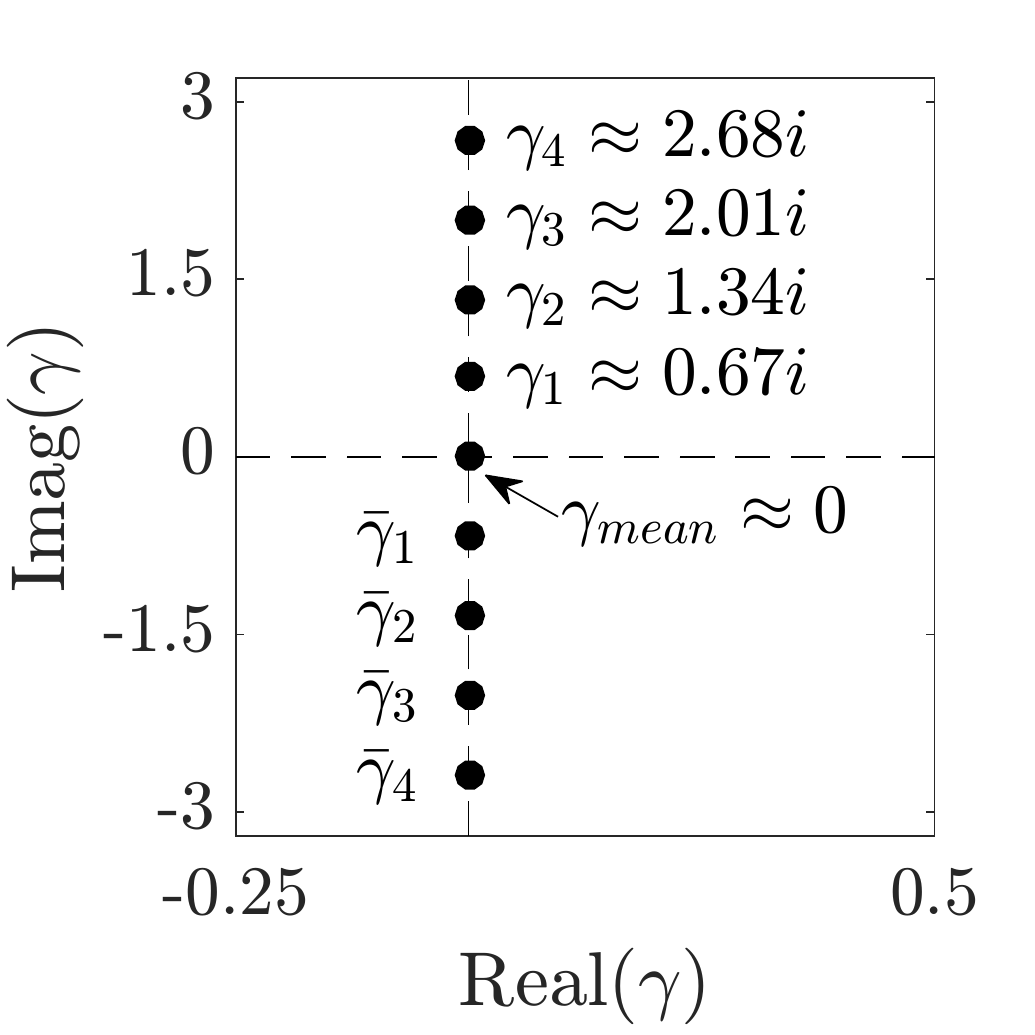}
	\end{subfigure}
	\caption{POD singular values $\sigma$ normalized by $\sigma_1$ (left) and DMD eigenvalues $\gamma$ (right) for limit-cycle flapping of a conventional flag with $Re =500, M_\rho = 0.1, K_B = 0.0001$.}
	\label{fig:singvals_convLC}
\end{figure}

\begin{figure}
%	\centering
	\begin{subfigure}[b]{0.245\textwidth}
        		\hspace*{0mm}
		\includegraphics[scale=0.27,trim={0cm 2.25cm 0cm 0cm},clip]{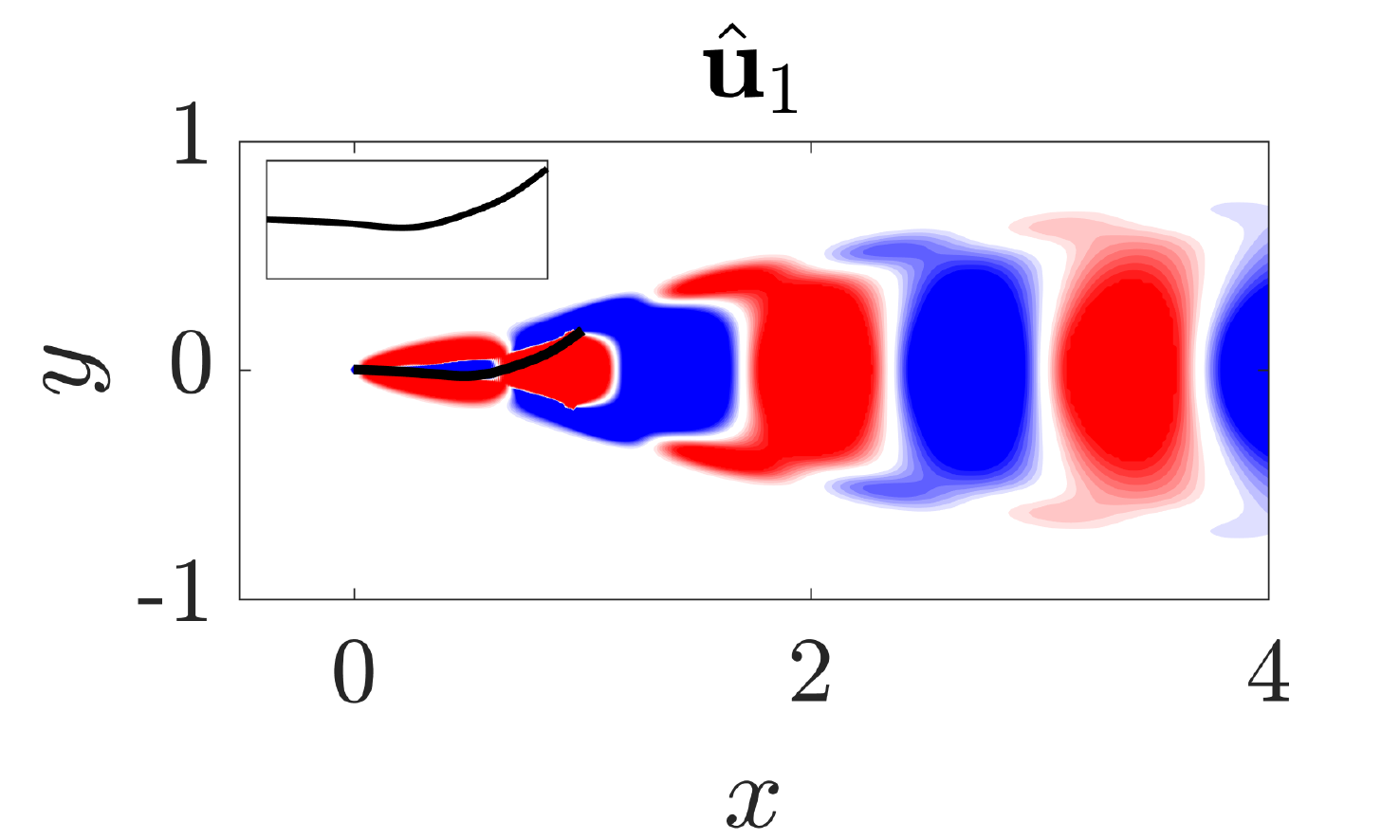}
	\end{subfigure}
	\begin{subfigure}[b]{0.245\textwidth}
		\hspace*{4.2mm}
        		\includegraphics[scale=0.27,trim={2.4cm 2.25cm 0cm 0cm},clip]{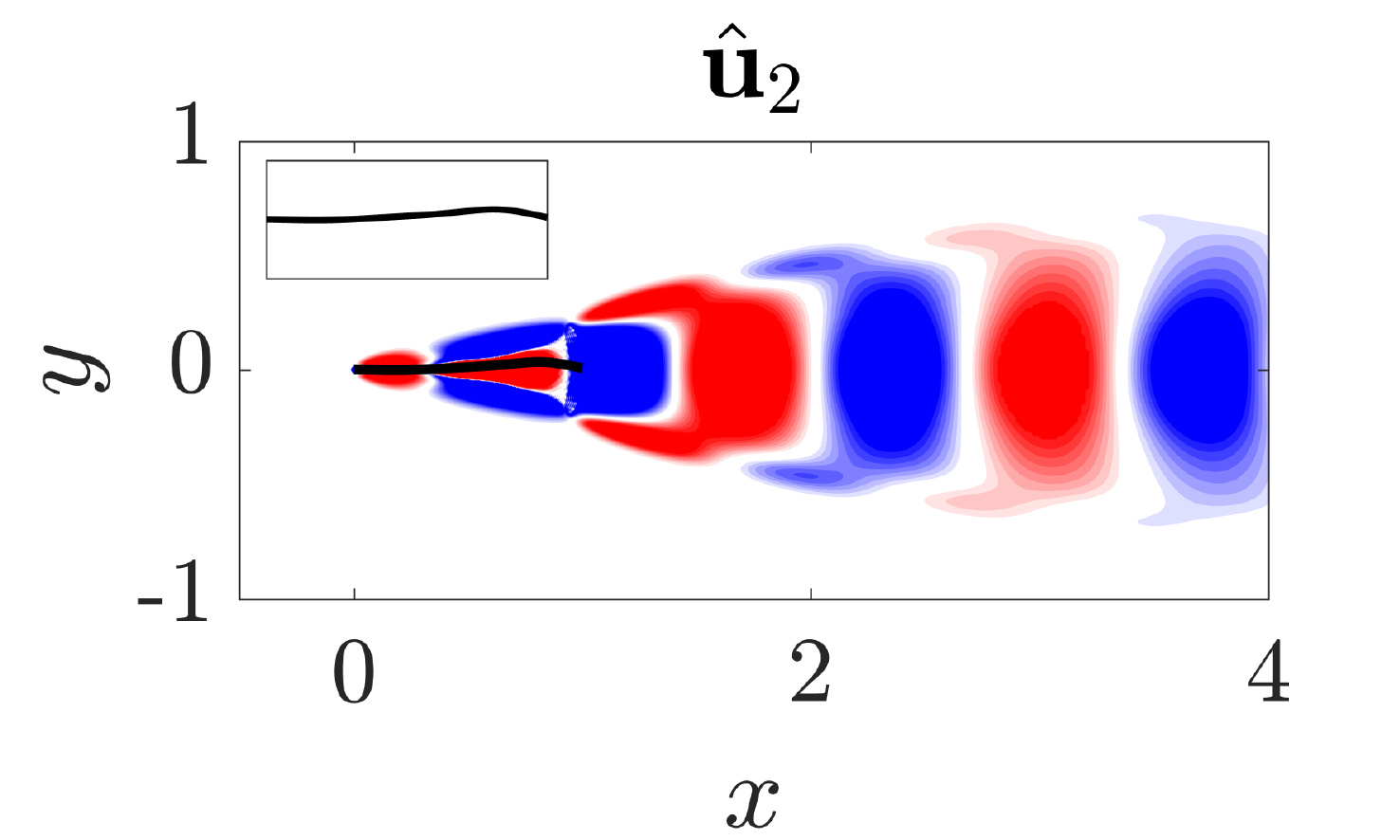}
	\end{subfigure}
    	\begin{subfigure}[b]{0.245\textwidth}
		\hspace*{2mm}
        		\includegraphics[scale=0.27,trim={2.4cm 2.25cm 0cm 0cm},clip]{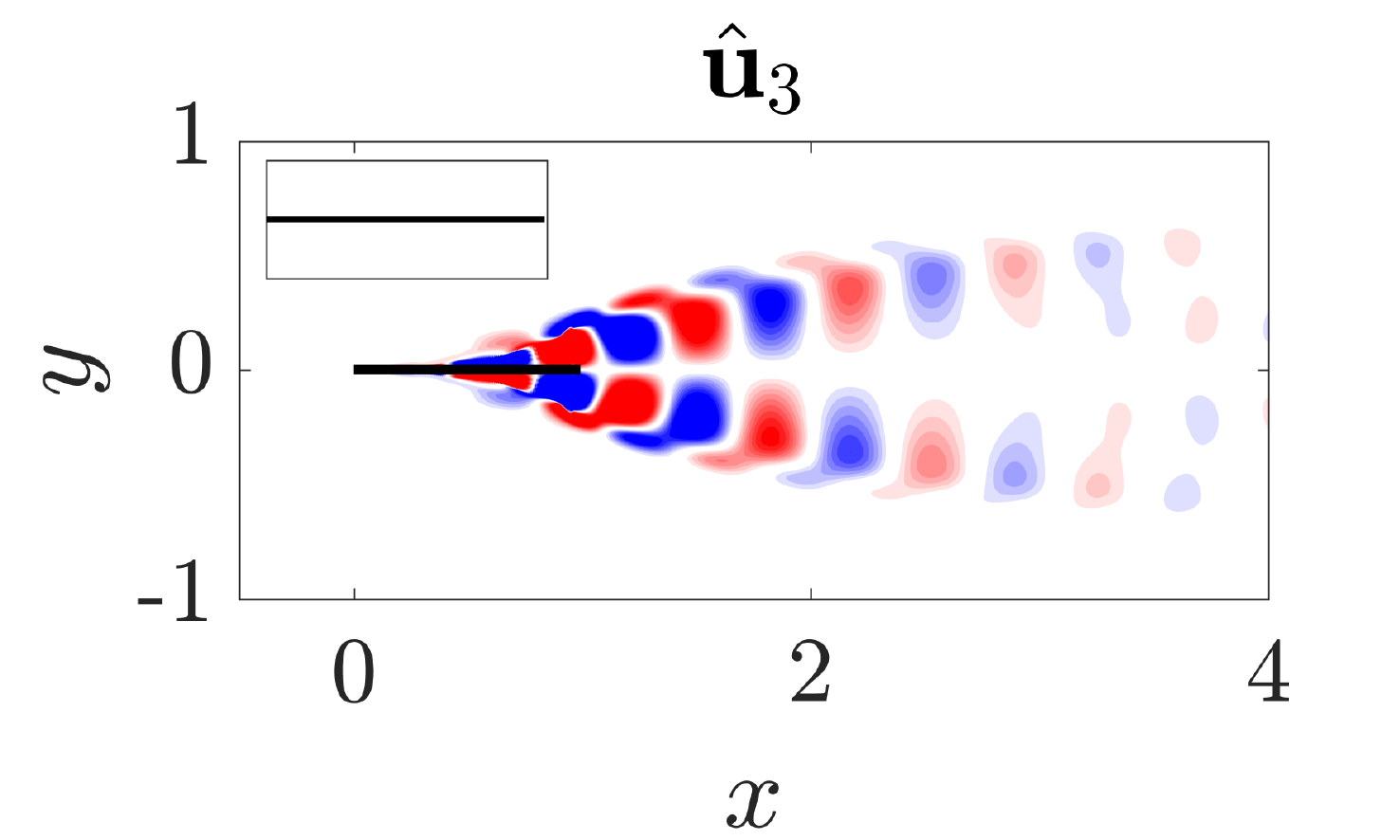}
	\end{subfigure}
	\begin{subfigure}[b]{0.245\textwidth}
		\hspace*{0mm}
        		\includegraphics[scale=0.27,trim={2.4cm 2.25cm 0cm 0cm},clip]{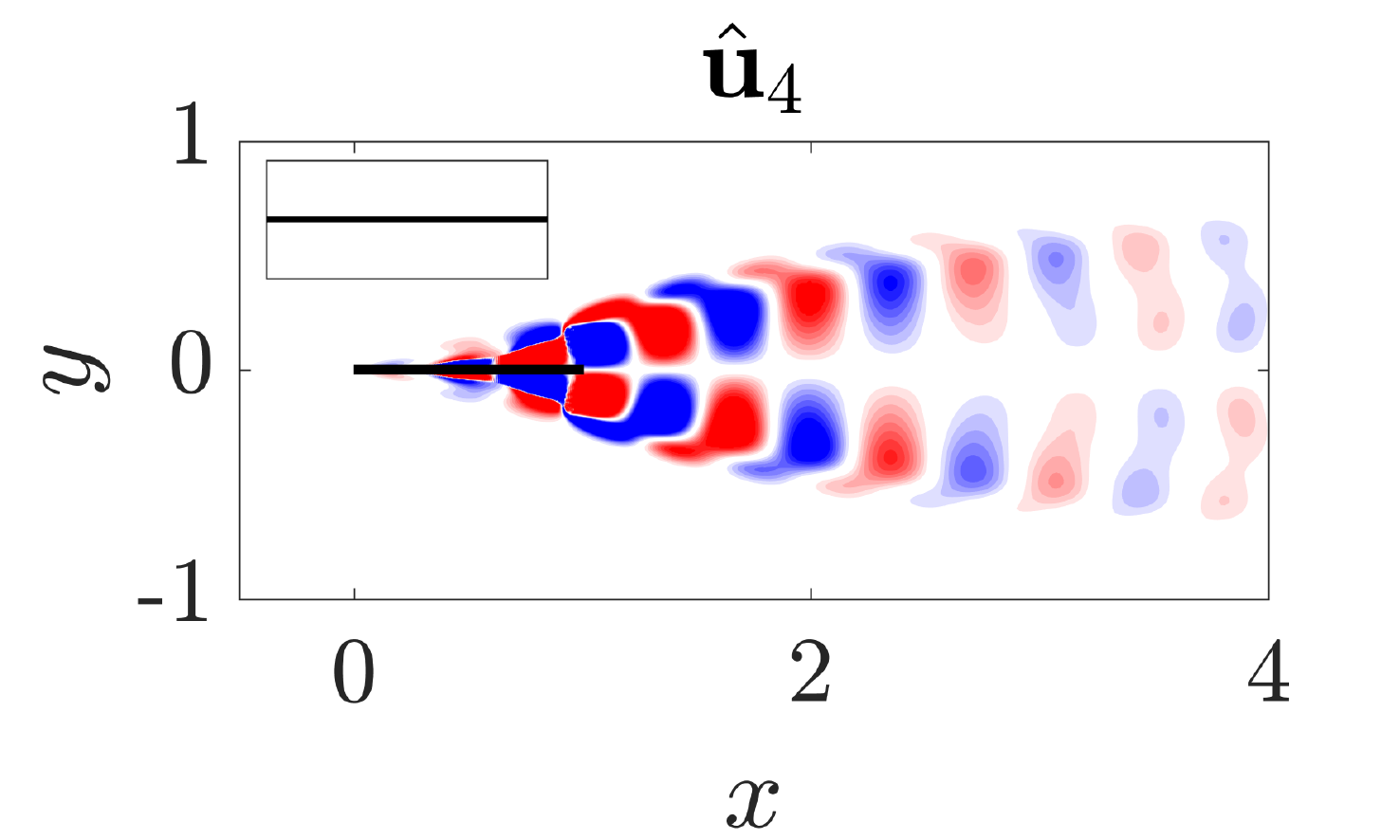}
	\end{subfigure}

	\begin{subfigure}[b]{0.245\textwidth}
        		\hspace*{0mm}
		\includegraphics[scale=0.27,trim={0cm 0cm 0cm 0cm},clip]{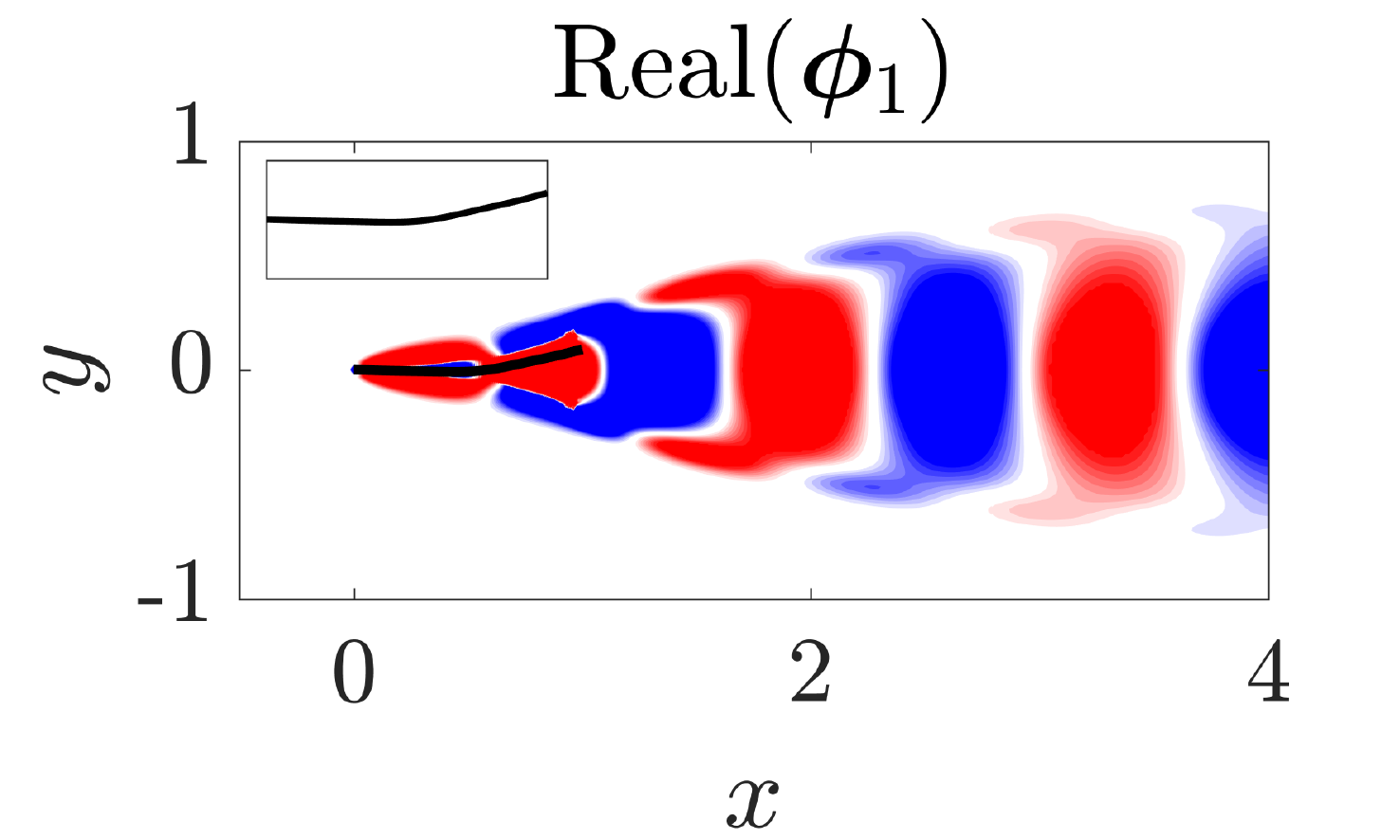}
	\end{subfigure}
	\begin{subfigure}[b]{0.245\textwidth}
		\hspace*{4.2mm}
        		\includegraphics[scale=0.27,trim={2.4cm 0cm 0cm 0cm},clip]{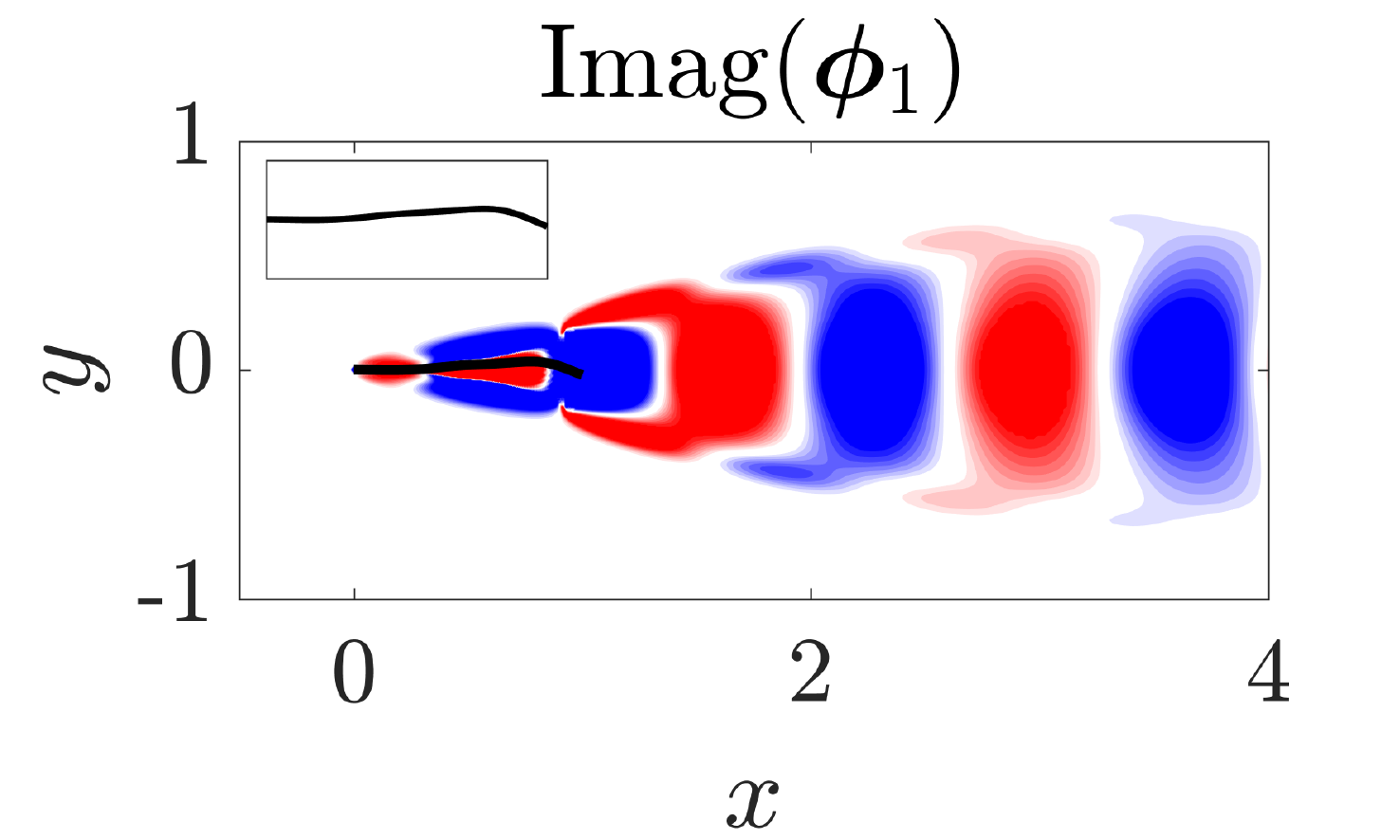}
	\end{subfigure}
    	\begin{subfigure}[b]{0.245\textwidth}
		\hspace*{2mm}
        		\includegraphics[scale=0.27,trim={2.4cm 0cm 0cm 0cm},clip]{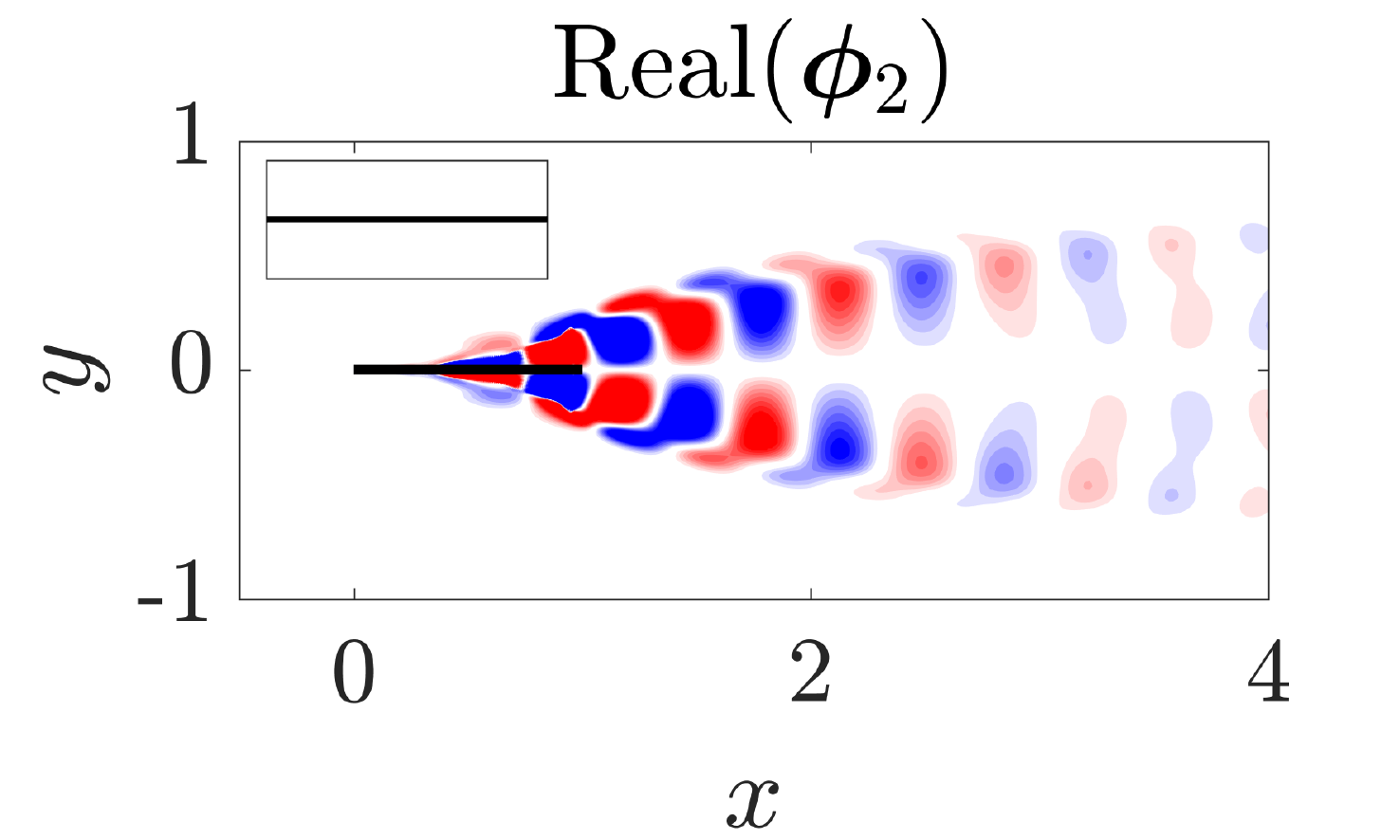}
	\end{subfigure}
	\begin{subfigure}[b]{0.245\textwidth}
		\hspace*{0mm}
        		\includegraphics[scale=0.27,trim={2.4cm 0cm 0cm 0cm},clip]{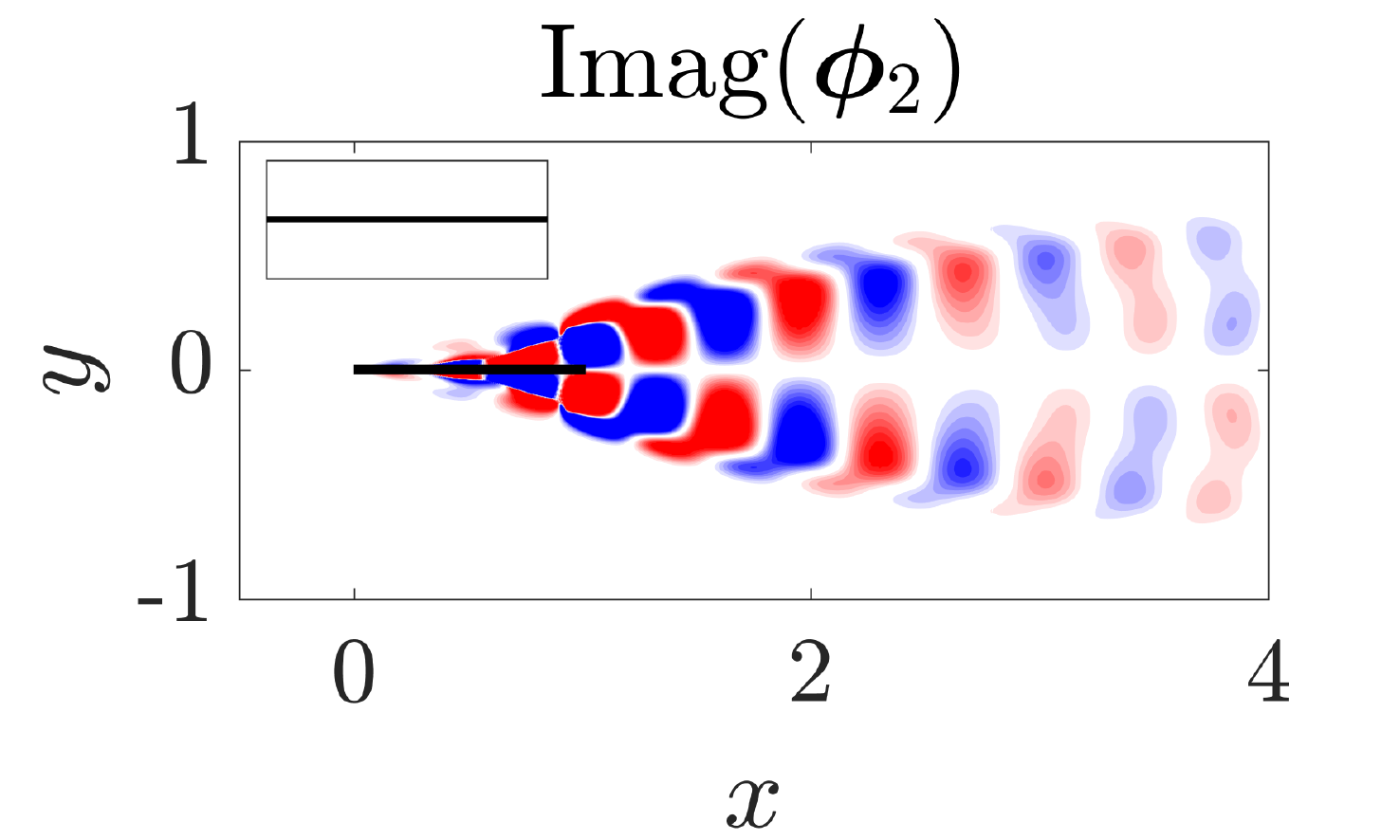}
	\end{subfigure}
    \caption{Leading POD (top row) and DMD (bottom row) modes for the limit-cycle conventional-flag problem.}
	\label{fig:mode_convLC}
\end{figure}

Figure \ref{fig:singvals_convLC} shows the singular values $\sigma$ from POD along with the DMD eigenvalues $\gamma$ of largest growth rate (real part). The four leading POD modes (which represent approximately $66\%$ of the total system energy) are shown in the top row of figure \ref{fig:mode_convLC}. Apart from the mode corresponding to the temporal mean, DMD modes typically come in complex conjugate pairs (\emph{e.g.}, the two leading modes are $\boldsymbol{\phi}_1, \bar{\boldsymbol{\phi}}_1$). We show in the bottom row of figure \ref{fig:mode_convLC} the real and imaginary parts of $\boldsymbol{\phi}_1$ and $\boldsymbol{\phi}_2$ (the mode corresponding to the temporal mean is not pictured). The POD and DMD modes are nearly identical since this system is characterized by a specific frequency (\emph{c.f.}, figure \ref{fig:conv_LC_tip}). The energetically optimal modes are therefore driving behavior at this dominant frequency and its harmonics. The flag behavior is conveyed through the leading two POD modes (leading complex-conjugate pair of DMD modes): these modes represent phase-shifted flapping at the dominant frequency to create the traveling-wave behavior of high spatial frequency that the flag undergoes for these parameters \citep{Connell2007}. The two leading POD modes (leading complex-conjugate pair of DMD modes) also demonstrate the creation and advection of vortices associated with flapping. Subsequent modes are not associated with flag flapping (the flag mode in the insert is undeformed), and instead describe the higher-harmonic response of the fluid to this dominant flapping motion.

\subsection{Chaotic flapping}
\label{sec:chaos}

Chaotic flapping of conventional flags can be triggered for flags of low stiffness ($K_B$) by increasing the flag mass ($M_\rho$). For flows at moderate Reynolds numbers of $O(1000)$, the system transitions with increasing mass from a stable equilibrium to limit-cycle flapping of increasing amplitude, then to chaotic flapping \citep{Connell2007}.  Similar transitions occur in inviscid fluids \citep{Alben2008}. We focus here on the case of moderate Reynolds number; establishing similarities in the driving mechanisms is an avenue of future work.

\begin{figure}
\centering
	\begin{subfigure}[b]{0.3\textwidth}
		\hspace*{-15mm}
        		\includegraphics[scale=0.25,trim={0cm 0cm 0cm 0cm},clip]{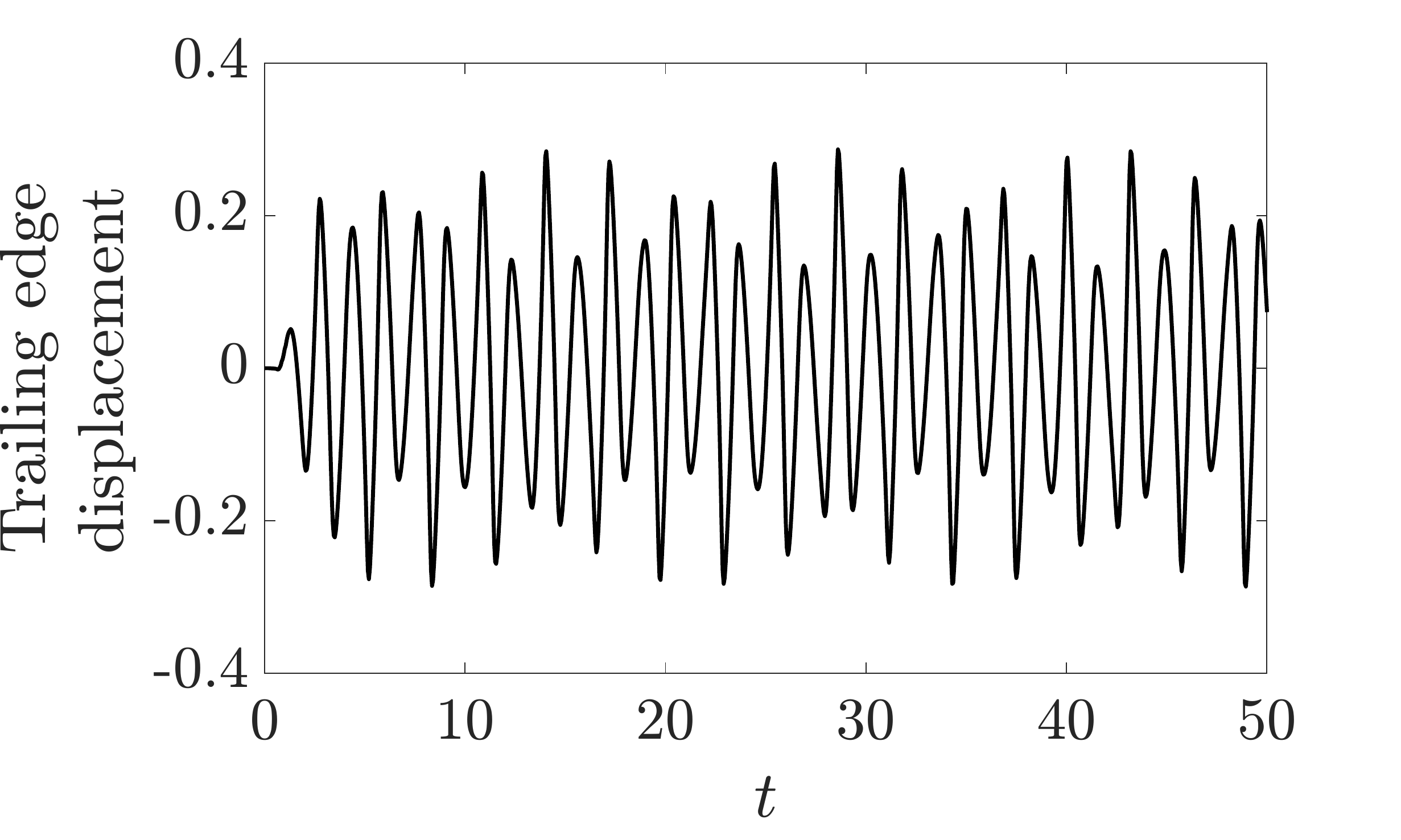}
	\end{subfigure}
	\begin{subfigure}[b]{0.3\textwidth}
		\hspace*{12mm}
        		\includegraphics[scale=0.25,trim={0cm 0cm 0cm 0cm},clip]{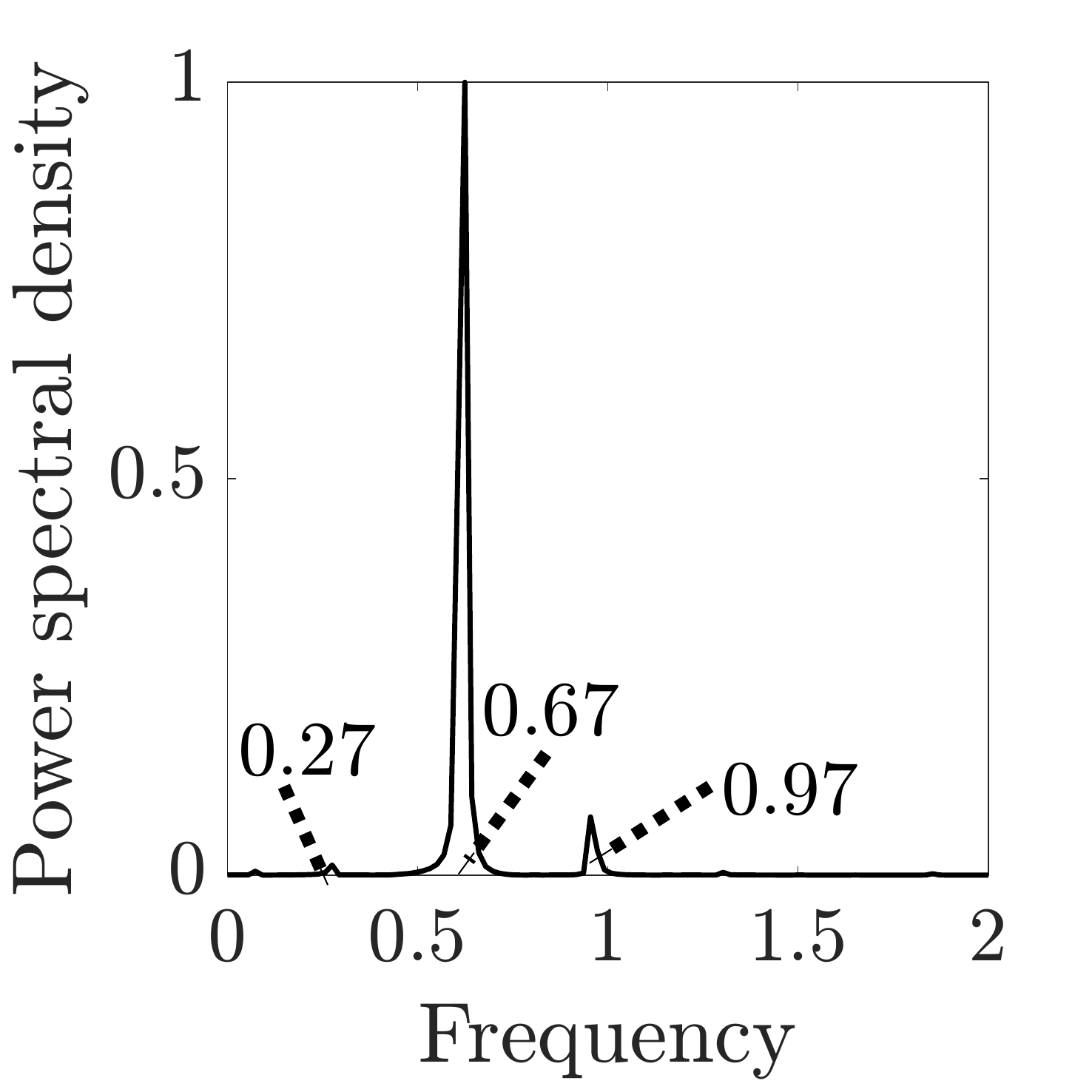}
	\end{subfigure}
	\caption{Transverse displacement (left) and spectral density (right) of the trailing edge of a flag in chaotic flapping for $Re =500,$ $M_\rho = 0.25$, and $K_B = 0.0001$.}
	\label{fig:conv_chaos_tip}
\end{figure}

\begin{figure}
	\begin{subfigure}[b]{0.245\textwidth}
        		\includegraphics[scale=0.35,trim={0cm 0cm 0cm 0cm},clip]{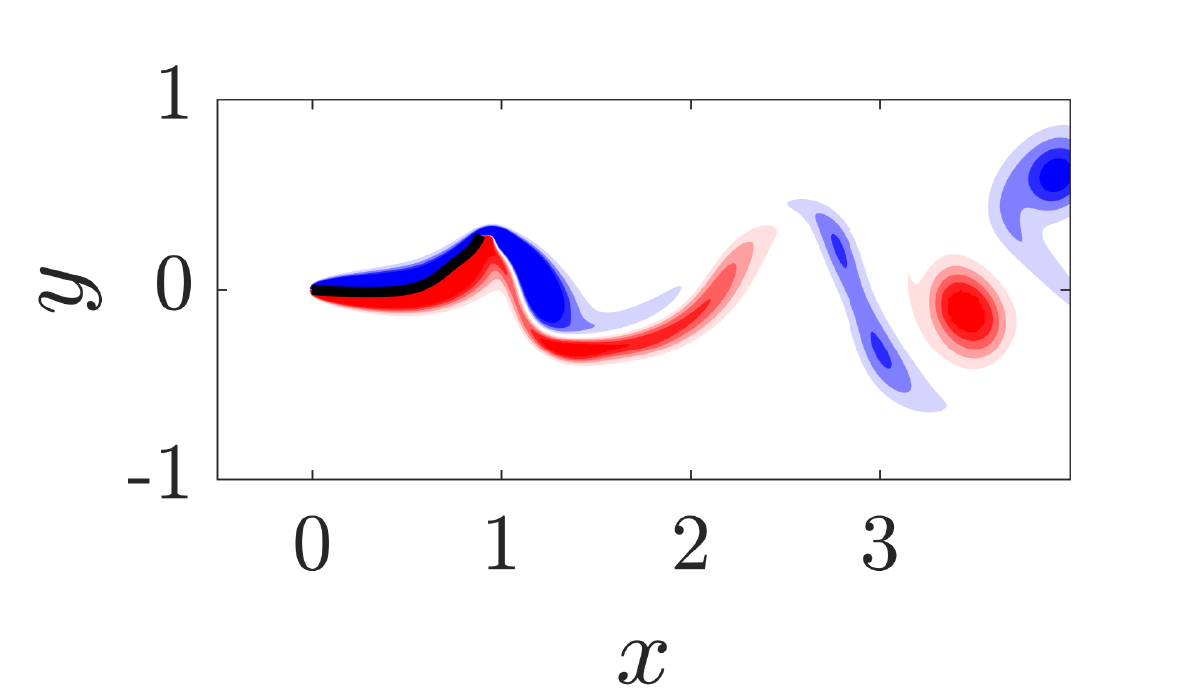}
	\end{subfigure}
	\begin{subfigure}[b]{0.245\textwidth}
		\hspace*{3.9mm}
        		\includegraphics[scale=0.35,trim={2cm 0cm 0cm 0cm},clip]{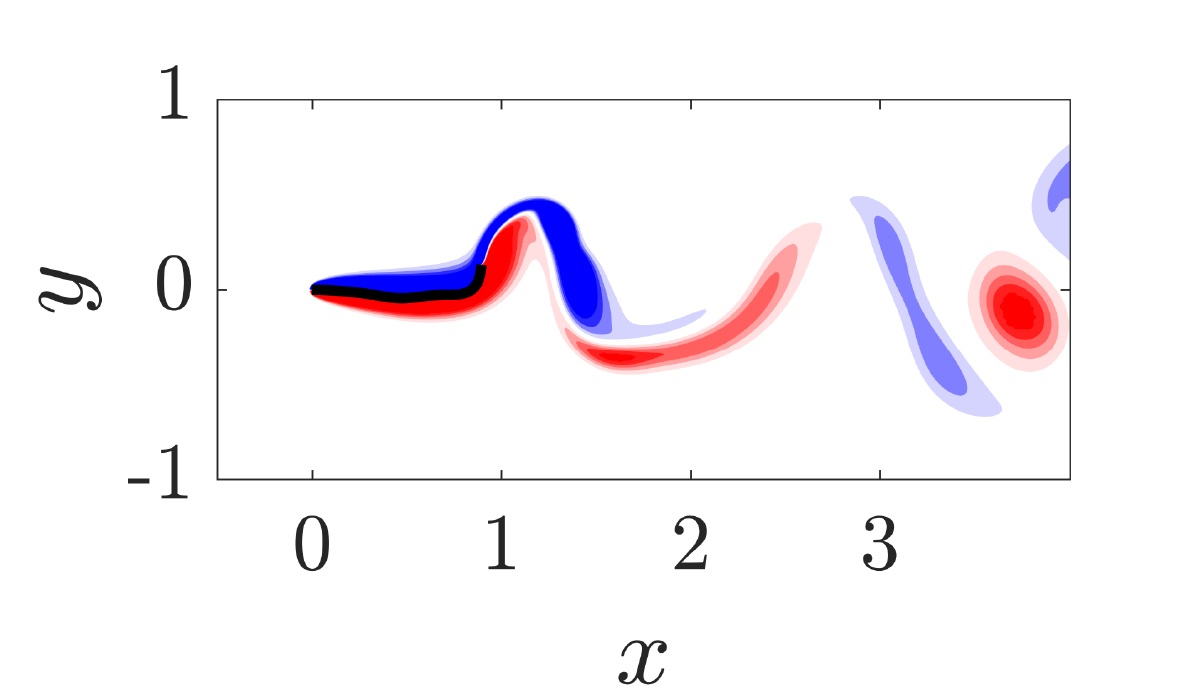}
	\end{subfigure}
    	\begin{subfigure}[b]{0.245\textwidth}
		\hspace*{1.5mm}
        		\includegraphics[scale=0.35,trim={2cm 0cm 0cm 0cm},clip]{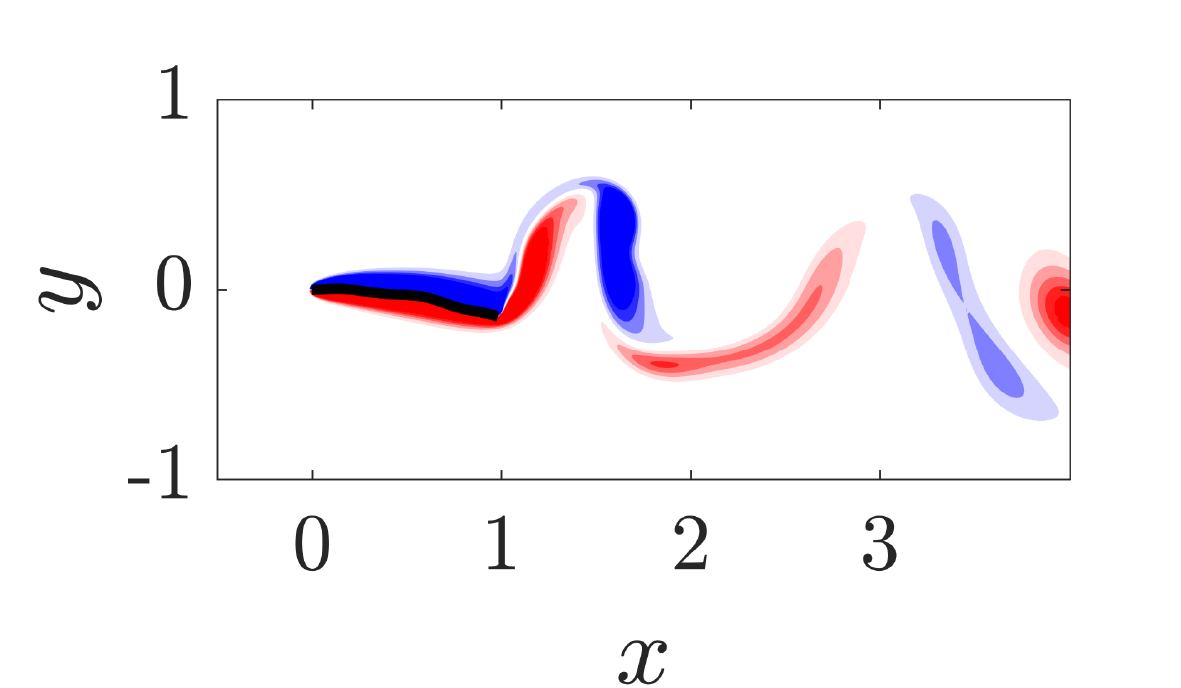}
	\end{subfigure}
	\begin{subfigure}[b]{0.245\textwidth}
		\hspace*{-0.8mm}
        		\includegraphics[scale=0.35,trim={2cm 0cm 0cm 0cm},clip]{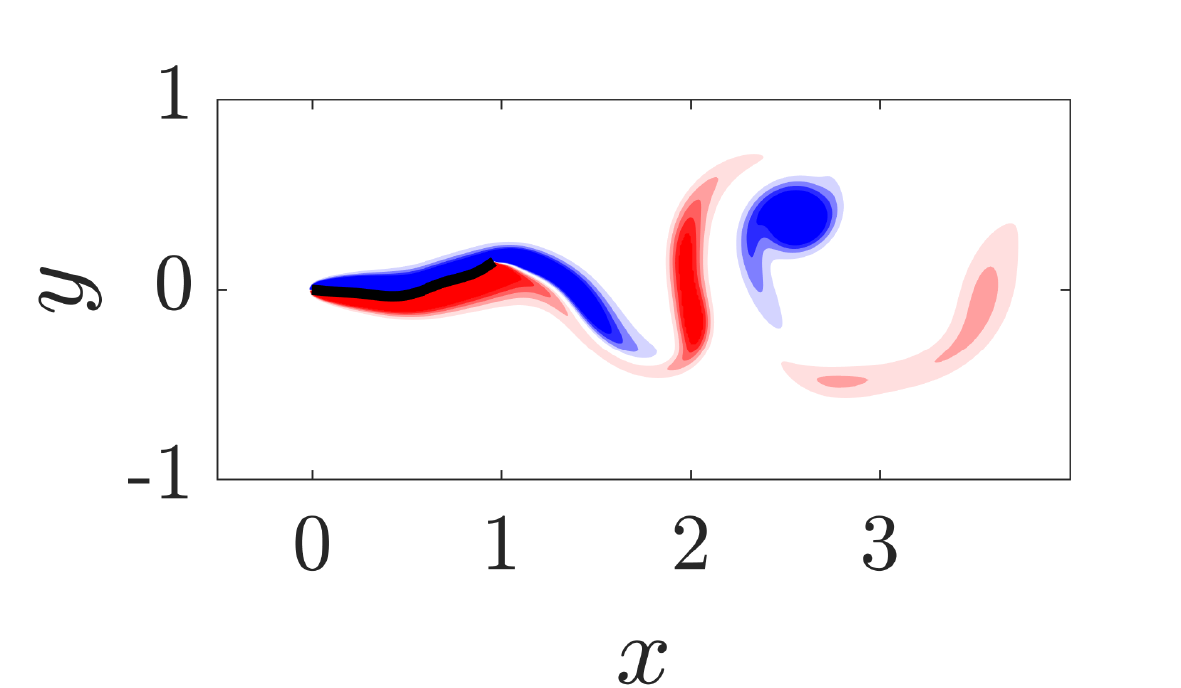}
	\end{subfigure}
    \caption{Snapshots for a flag in chaotic flapping with $Re =500, M_\rho = 0.25, K_B = 0.0001$. Contours are of vorticity, in 18 increments from -5 to 5.}
	\label{fig:chaos_snaps}
\end{figure}

We investigate the route to chaotic flapping here by choosing $M_\rho = 0.25$, which is near the critical value where the system transitions from limit-cycle flapping. The trailing-edge displacement and corresponding spectral density for this regime are shown in figure \ref{fig:conv_chaos_tip}. We also show in figure \ref{fig:chaos_snaps} snapshots of the system over $t\in[28.6, 30.2]$. Note the increase in flapping amplitude compared with the $M_\rho = 0.18$ case described above (\emph{c.f.}, figure \ref{fig:conv_LC_tip}). Moreover, in chaotic flapping there are multiple frequencies present at non-integer harmonics of the dominant frequency. These non-integer frequencies were first observed by \citet{Connell2007}, and the mechanism that introduces them remains unexplained. 

Using DMD within our FSI framework, we propose a mechanism in which chaotic flapping is instigated by the increase in flapping amplitude associated with the increased mass ratio. This increase in amplitude leads the flag to become sufficiently bluff to the flow at its peak deflection that a bluff-body wake instability arises and interacts triadically with the dominant flapping behavior to produce the subdominant flapping frequencies observed in figure \ref{fig:conv_chaos_tip}. DMD is selected here to isolate behavior at distinct frequencies. This can be done in a POD context using spectral POD (SPOD) \citep{Towne2017}, and future work could compare the results between DMD and SPOD.

\begin{figure}
\centering
	\begin{subfigure}[b]{0.45\textwidth}
		\hspace*{3mm}
            	\includegraphics[scale=0.45,trim={0cm 0cm 0cm 0cm},clip]{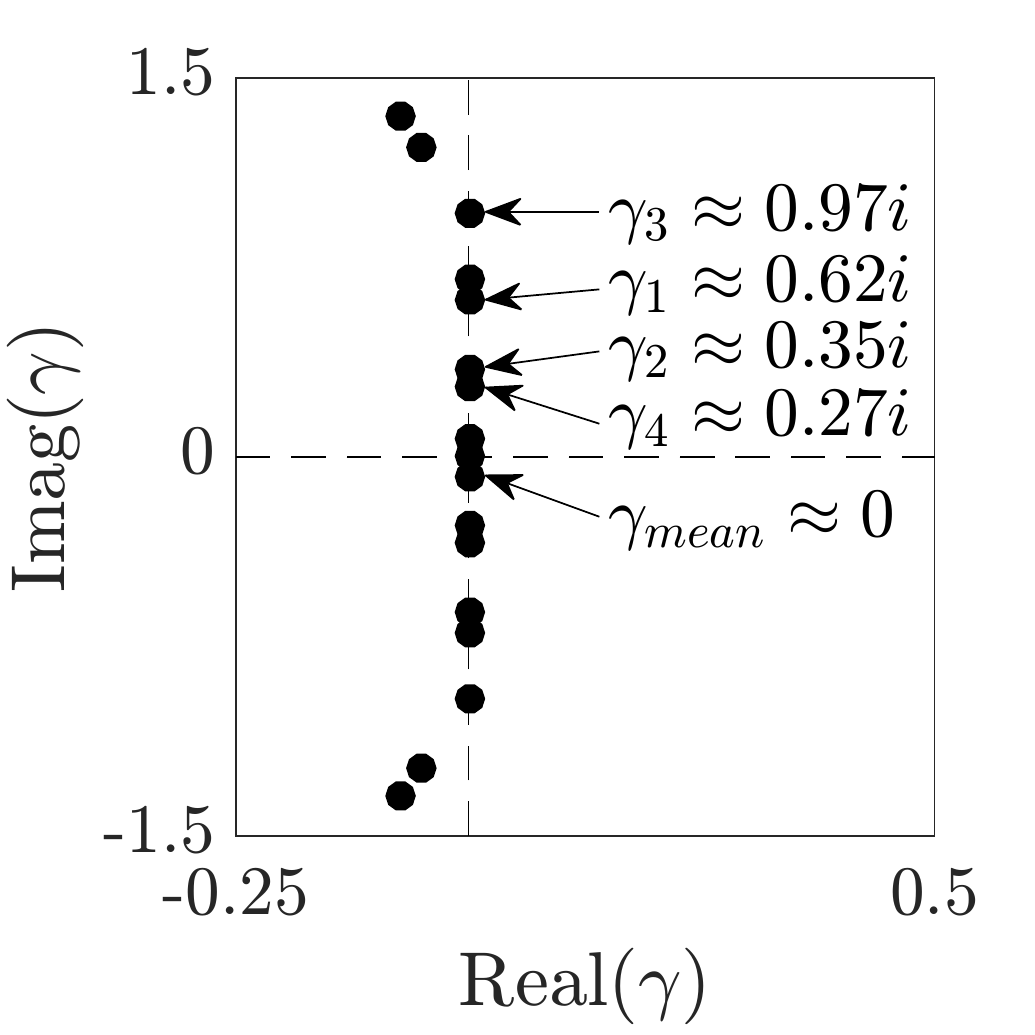}
	\end{subfigure}
	\caption{DMD eigenvalues $\gamma$ for chaotic flapping of a conventional flag with $Re =500, M_\rho = 0.25, K_B = 0.0001$.}
	\label{fig:evals_chaos}
\end{figure}

\begin{figure}
	\begin{subfigure}[b]{0.245\textwidth}
        		\hspace*{0mm}
		\includegraphics[scale=0.27,trim={0cm 2.25cm 0cm 0cm},clip]{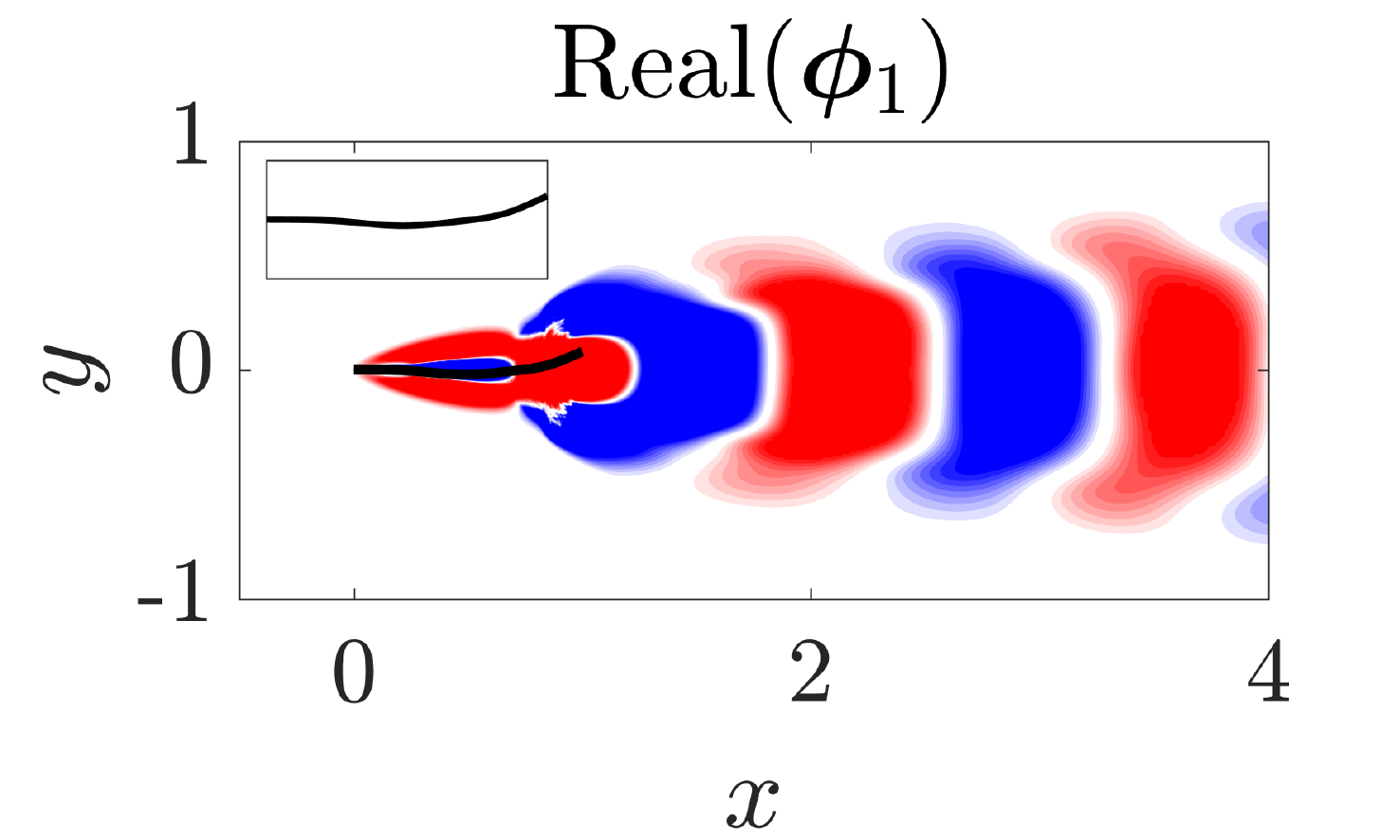}
	\end{subfigure}
	\begin{subfigure}[b]{0.245\textwidth}
		\hspace*{4.2mm}
        		\includegraphics[scale=0.27,trim={2.4cm 2.25cm 0cm 0cm},clip]{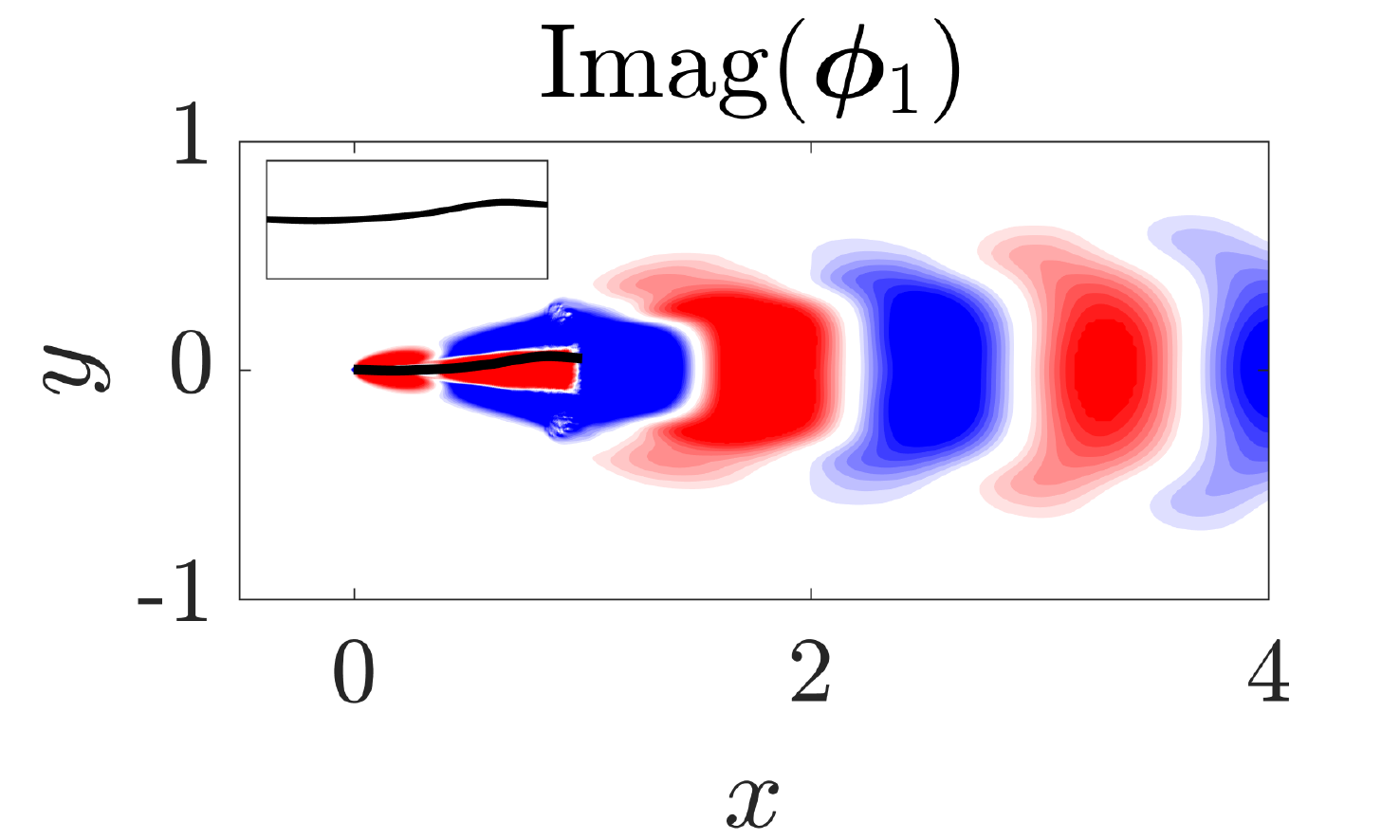}
	\end{subfigure}
    	\begin{subfigure}[b]{0.245\textwidth}
		\hspace*{2mm}
        		\includegraphics[scale=0.27,trim={2.4cm 2.25cm 0cm 0cm},clip]{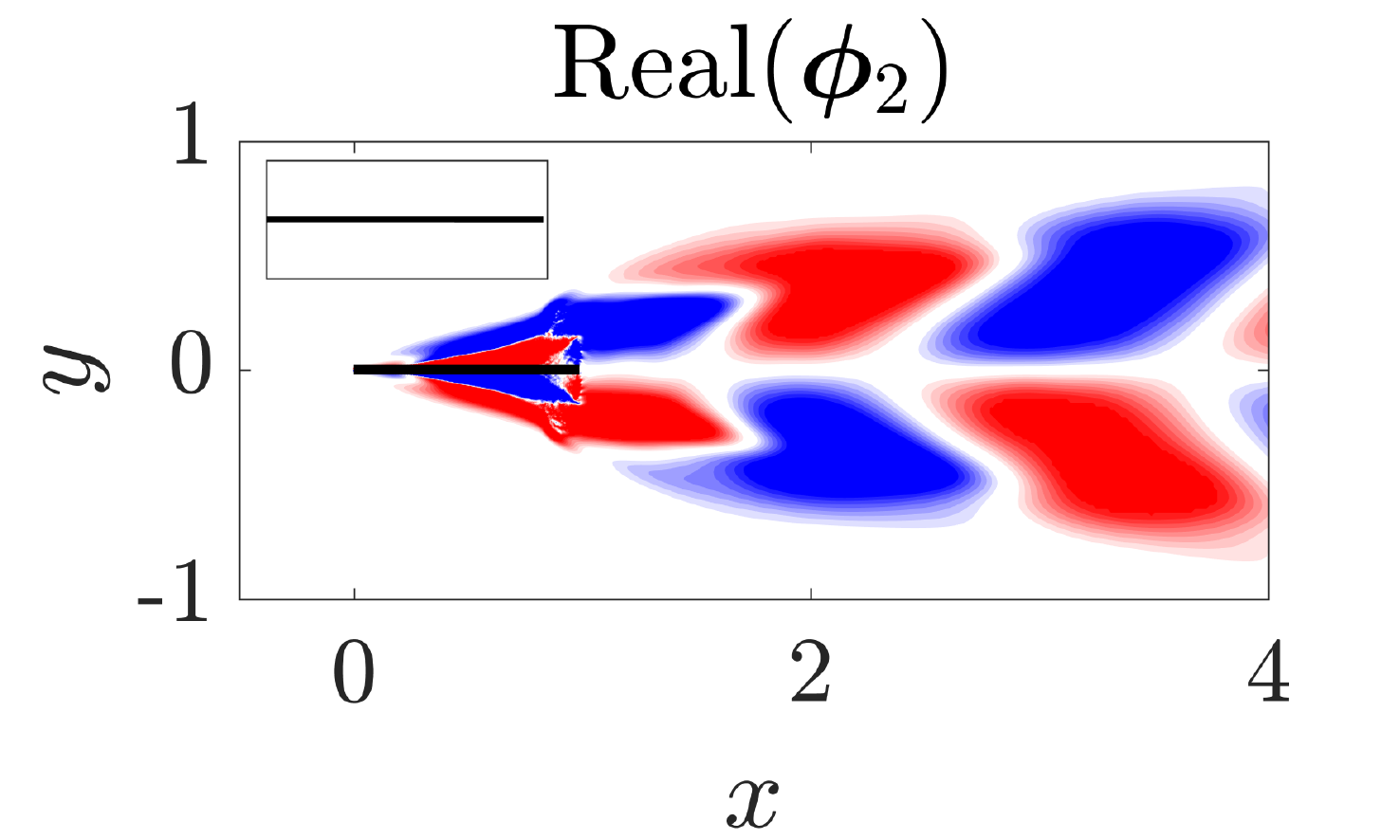}
	\end{subfigure}
	\begin{subfigure}[b]{0.245\textwidth}
		\hspace*{0mm}
        		\includegraphics[scale=0.27,trim={2.4cm 2.25cm 0cm 0cm},clip]{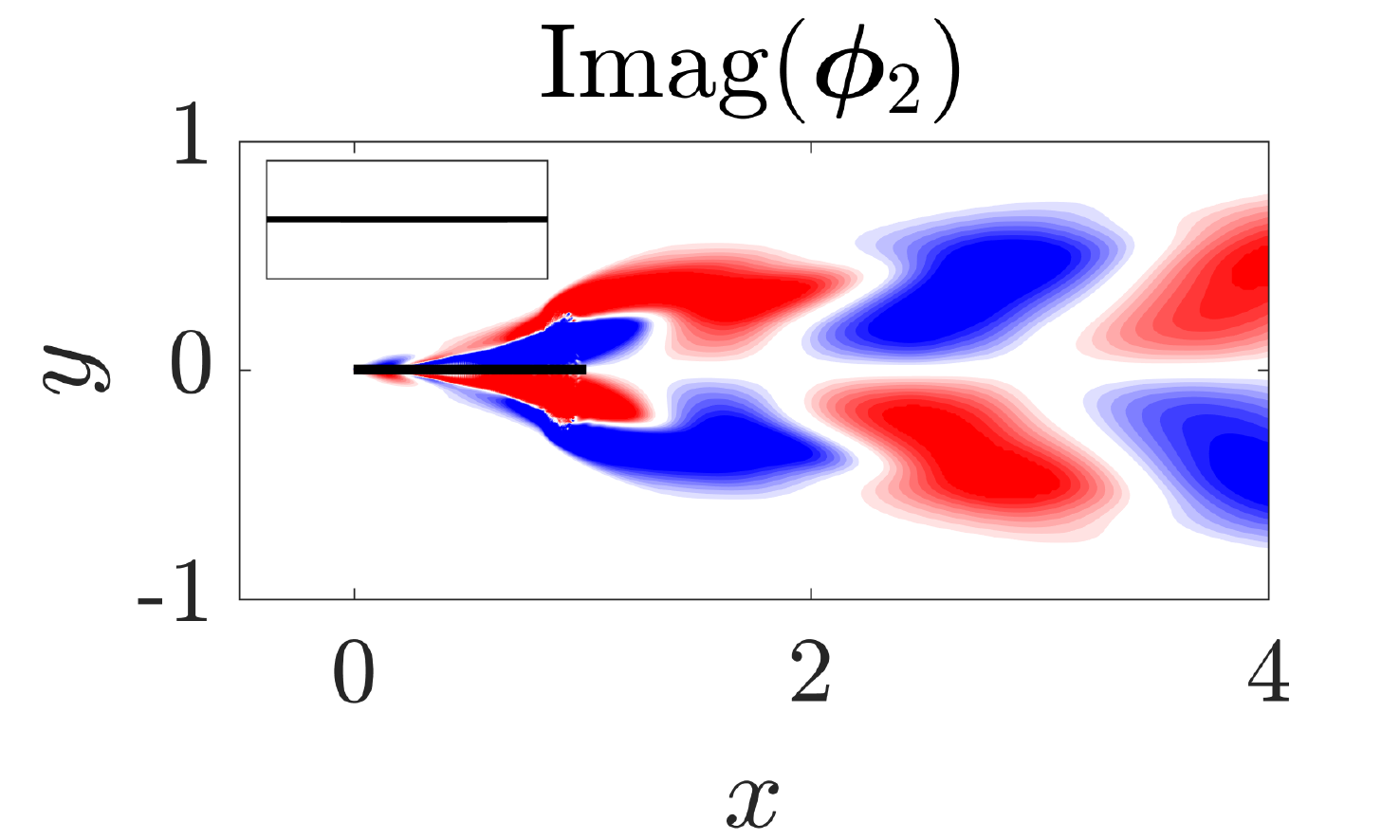}
	\end{subfigure}
	
	\begin{subfigure}[b]{0.245\textwidth}
        		\hspace*{0mm}
		\includegraphics[scale=0.27,trim={0cm 0cm 0cm 0cm},clip]{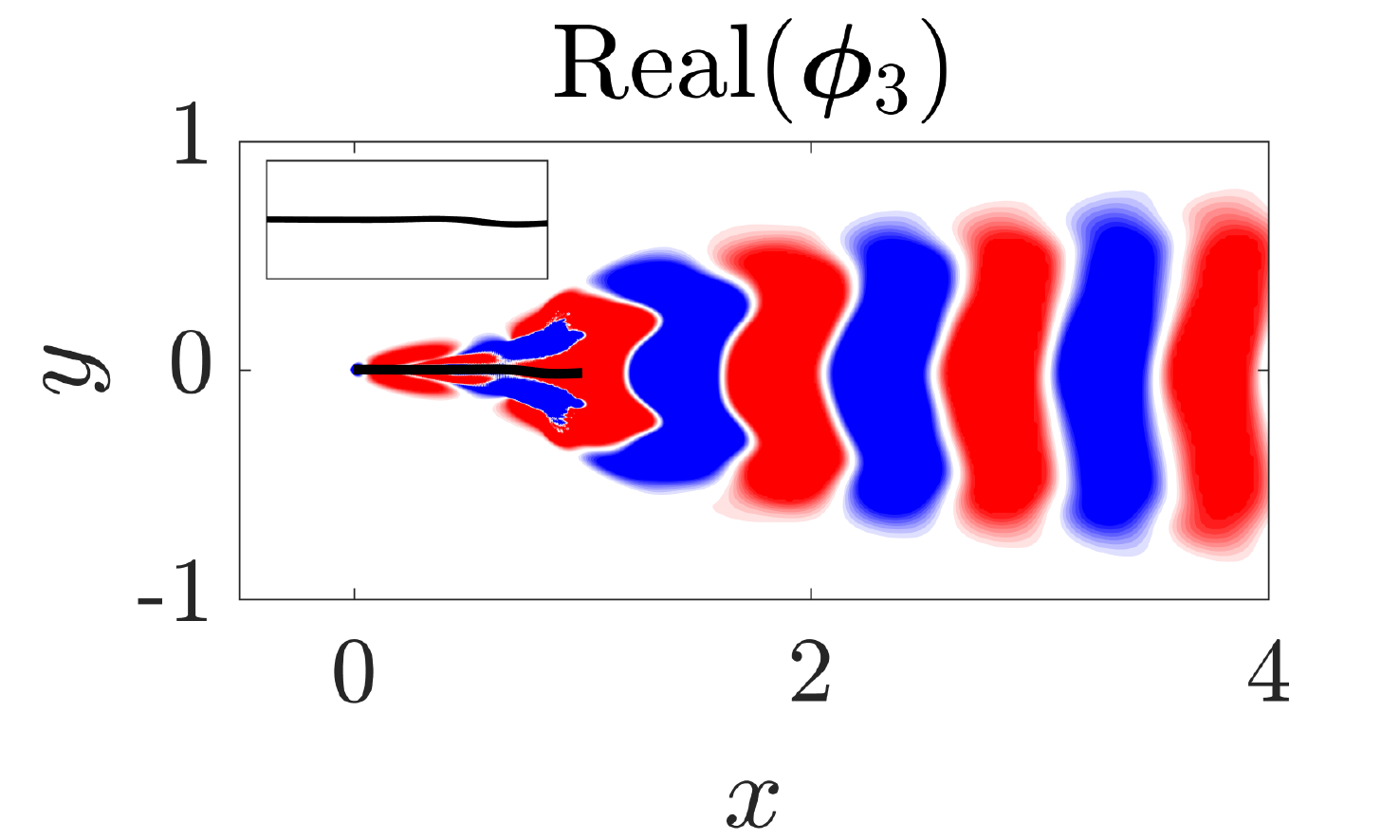}
	\end{subfigure}
	\begin{subfigure}[b]{0.245\textwidth}
		\hspace*{4.2mm}
        		\includegraphics[scale=0.27,trim={2.4cm 0cm 0cm 0cm},clip]{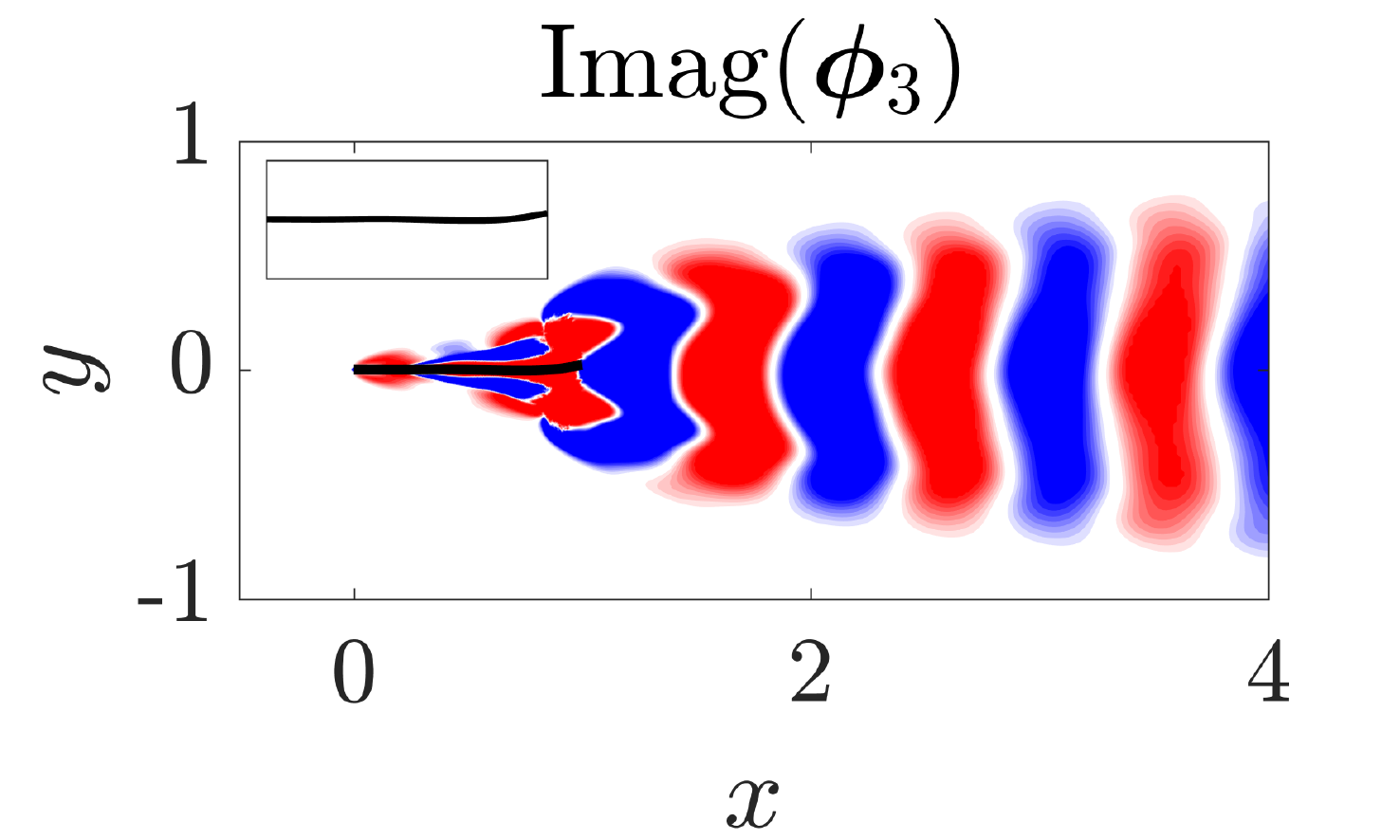}
	\end{subfigure}
    	\begin{subfigure}[b]{0.245\textwidth}
		\hspace*{2mm}
        		\includegraphics[scale=0.27,trim={2.4cm 0cm 0cm 0cm},clip]{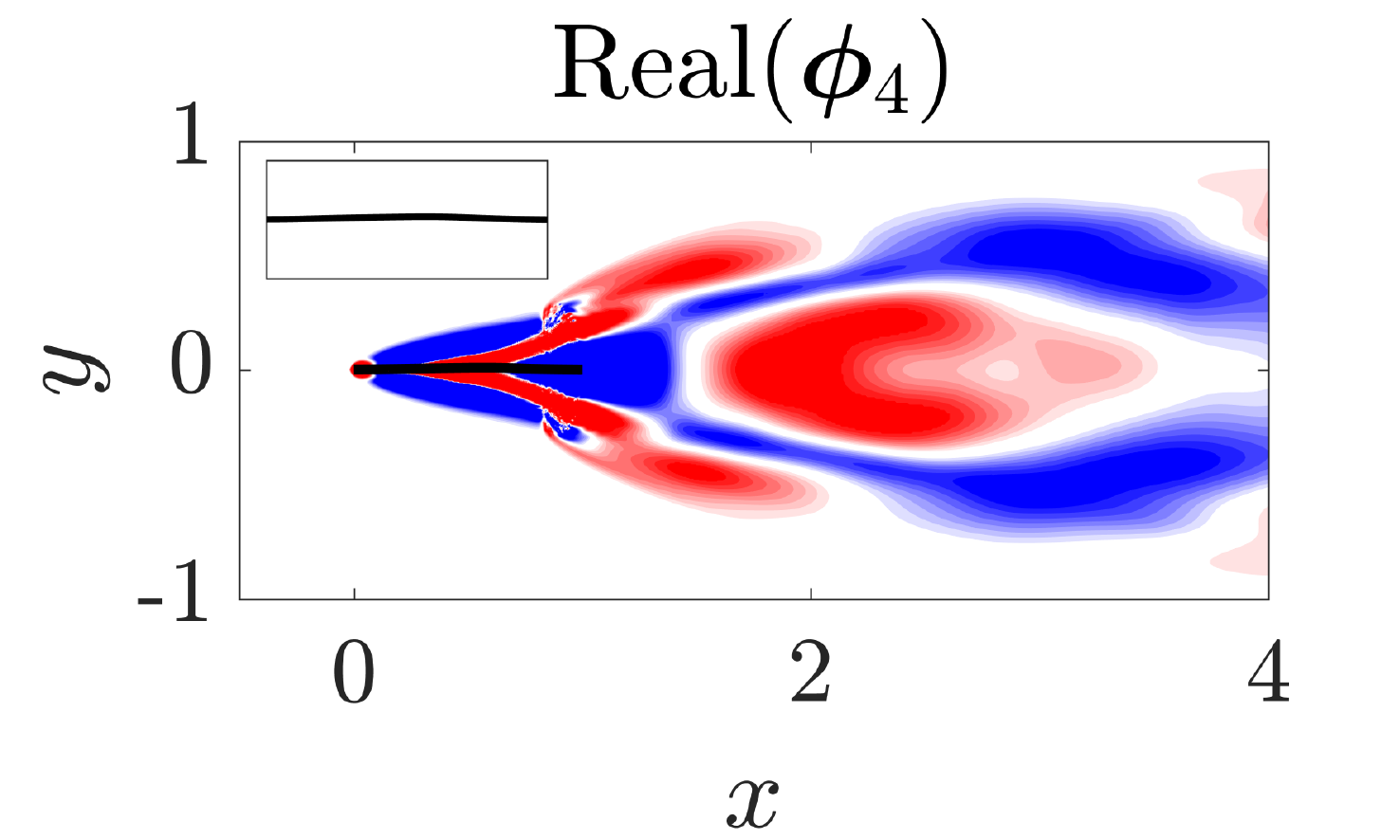}
	\end{subfigure}
	\begin{subfigure}[b]{0.245\textwidth}
		\hspace*{0mm}
        		\includegraphics[scale=0.27,trim={2.4cm 0cm 0cm 0cm},clip]{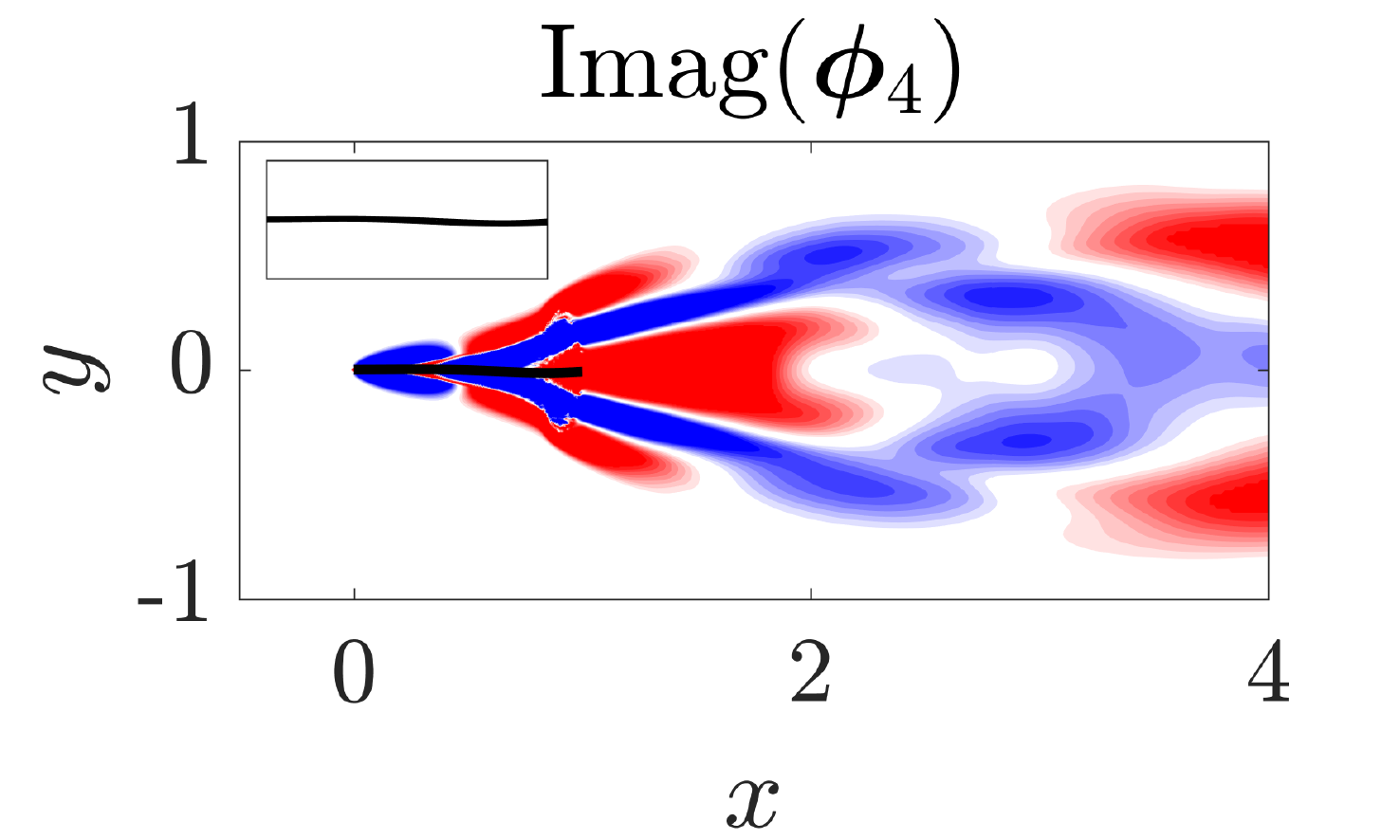}
	\end{subfigure}
	    \caption{Leading DMD modes for chaotic flapping of a conventional flag with $Re = 500, M_\rho =0.25, K_B=0.0001$.}
	\label{fig:mode_convchaos}
\end{figure}

The DMD eigenvalues $\gamma$ and four leading modes ${\boldsymbol{\phi}}$ (omitting the mode associated with the mean) for the chaotic case of $M_\rho = 0.25$ are shown in figures \ref{fig:evals_chaos} and \ref{fig:mode_convchaos}. The dominant and non-integer harmonic frequencies from the spectral density plot of figure \ref{fig:conv_chaos_tip} manifest themselves in DMD modes $\boldsymbol{\phi}_1,$ $\boldsymbol{\phi}_3$, and $\boldsymbol{\phi}_4$ (see the corresponding eigenvalues in figure \ref{fig:evals_chaos}). Note that despite the significant change in behavior from the limit-cycle regime, $\boldsymbol{\phi}_1$ remains largely unchanged. Yet, due to the increased system complexity, flapping is no longer conveyed entirely through the first mode, and both $\boldsymbol{\phi}_3$ and $\boldsymbol{\phi}_4$ are associated with flapping motion and a correlated set of flow features.

By contrast, $\boldsymbol{\phi}_2$ is not associated with flapping (the flag mode in the insert is undeformed). This is consistent with the absence of the $\gamma_2$ frequency in the spectral density plot of figure \ref{fig:conv_chaos_tip}. Thus, the mode represents a response of the fluid to the dominant flapping motion. The pronounced shear layers at the top and bottom peak displacement and the corresponding wake vortices are reflective of a bluff-body vortex-shedding mode that appears because of the increased flapping amplitude compared with the limit-cycle case. This is further evidenced by the modal frequency, which agrees with the classical 0.2 Strouhal scaling \citep{Roshko1954} when normalized by the projected length of the maximum peak-to-peak-amplitude ($0.35 \times 0.5 \approx 0.18$). Note also that $\gamma_2$ is not a sub-harmonic of the dominant flapping frequency $\gamma_1$, and thus this bluff-body mode is reflective of the appearance of a new physical mechanism rather than of resonance or harmonic interactions.

This bluff-body mode is key to understanding the sub-dominant flapping behavior of the flag: the sub-dominant frequencies seen in figure \ref{fig:conv_chaos_tip} arise as triadic combinations of the frequencies of the dominant flapping mode and the bluff-body mode; \emph{i.e.}, $\gamma_3 = \gamma_1 + \gamma_2$ and $\gamma_4 = \gamma_1 - \gamma_2$. These triadic interactions are necessitated by the quadratic nonlinearity of the advective term in the Navier-Stokes equations.

\section{Conclusions}

We presented a formulation for performing data analysis on FSI problems that accounts for both the fluid and the structure. We designed this formulation to be compatible with the manner in which data is typically obtained for experiments and nonconforming mesh simulations. As part of this framework, we defined a physically meaningful norm for FSI systems. We  considered POD and DMD because of their widespread use, but extensions to other methods are straightforward. 

Our formulation was first applied to limit-cycle flag flapping. Because of the dominant frequency associated with this limit-cycle behavior, both POD and DMD give similar decompositions. The leading two POD modes (leading complex-conjugate pair of DMD modes) convey both the flapping information of the flag and the dominant vortical structures associated with this motion. Subsequent modes describe harmonic responses in the fluid to the flapping described in the leading modes. 

Next, the physical mechanism driving chaotic flapping was clarified. \citet{Connell2007} identified that the transition from limit-cycle flapping to chaotic flapping coincides with the appearance of a new flapping frequency near the 3/2 harmonic of the dominant flapping frequency. We identified the mechanism driving this non-integer harmonic through a DMD analysis. We first demonstrated that at the onset of chaos, the flag becomes sufficiently bluff at its peak deflection to initiate a bluff-body wake instability. This is in contrast to limit-cycle flapping, where flapping amplitudes are smaller and this bluff-body instability is not instigated. The associated shedding frequency of this new behavior coincides with the Strouhal scaling of 0.2 common to bluff-body flows \citep{Roshko1954}. Moreover, we demonstrated that this bluff-body mode combines triadically with the dominant flapping behavior to produce the observed flapping near the 3/2 harmonic (and the other sub-dominant flapping frequencies). 

Finally, we note that data analysis is often used to develop reduced-order models of complex flow. For FSI systems, these models are typically derived by performing a data-driven decomposition of the fluid and coupling this to the full governing equations for the structure (see \citet{Dowell2001} for a review). This approach may require more modes than those derived from a combined fluid-structure treatment, and there are avenues for future work in evaluating the efficiency of our proposed data-analysis technique in the context of reduced-order models.

\section{Acknowledgments}

AJG and TC acknowledge funding through the BOSCH Bern program and through the AFOSR (grant number FA9550-14-1-0328). AJG is also grateful to Dr. Scott Dawson for his thoughtful comments on an early version of the manuscript.

\bibliographystyle{jfm}
\bibliography{data}{}

\end{document}